\documentclass[epj,twocolumn]{webofc}
\usepackage[varg]{txfonts}   

\usepackage{color}
\usepackage{graphicx} 
\usepackage{amssymb,amsmath,enumerate} 
\begin{document}

\title{Spectrum and Structure of Excited Baryons with CLAS}

\author{\firstname{Volker~D.} \lastname{Burkert}\inst{1}\fnsep\thanks{Talk presented at the CRC-16 Symposium, Bonn University, June 6-9, 2016}, for the CLAS collaboration
}

\institute{Thomas Jefferson National Accelerator Facility \\ 12000 Jefferson Avenue, Newport News, Virginia, USA  } 
\abstract{%
  In this contribution I discuss recent results in light quark baryon spectroscopy involving CLAS data and higher level analysis
  results from the partial wave analysis by the Bonn-Gatchina group. New baryon states were discovered largely
  based on the open strangeness production channels $\gamma p \to K^+ \Lambda$ and $\gamma p \to K^+ \Sigma^0$. 
  The data illustrate the great potential of the kaon-hyperon channel in the discovery of higher mass baryon resonances 
  in s-channel production. Other channels with discovery potential, such as $\gamma p \to p \omega$ and $\gamma p \to 
  \phi p$ are also discussed. In the second part I will demonstrate on data the sensitivity of meson electroproduction to 
  expose the active degrees of freedom underlying resonance transitions as a function of the probed distance scale. 
  For several of the prominent excited states in the lower mass range the short distance behavior is described by a
  core of three dressed-quarks with running quark mass, and meson-baryon contributions make up significant parts of
  the excitation strength at large distances. Finally, I give an outlook of baryon resonance physics at the 12 GeV CEBAF 
  electron accelerator. 
}
\maketitle

\section{Foreword}

In the first physics talk at this symposium Ulrike Thoma asked me to present the research on excited baryon states with the CLAS detector at Jefferson Lab.  Before I come to this subject I like to point out how pleased I am to participate in and to contribute to the celebration of the CRC 16 program, whose impact on baryon spectroscopy is well known in the broader hadron community. 
Many of the results I show in my presentation would not have been possible without the CRC 16 project and its phenomenological component, which is generally known as the Bonn-Gatchina (BnGa) coupled-channel analysis framework.
When I refer to newly discovered baryon states they will have data from CLAS as a major input, but the high level analysis includes other data sets as well and the results of this analysis is the responsibility of the CRC 16 project, which by having strong experimental and analysis components is making lasting contributions to our understanding of the nucleon spectrum.
\begin{figure}[tb]

\hspace{-0.4cm}\includegraphics[width=8.8cm,height=7.0cm,clip]{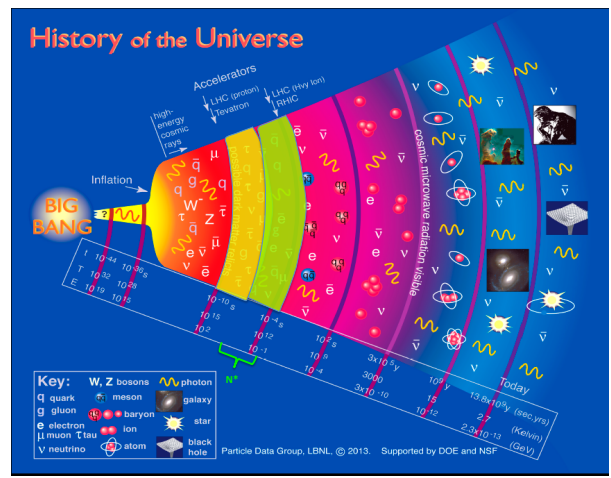}
\caption{The evolution of the universe as pictured by the LBNL Particle Data Group, 2013. 
The area highlighted in yellow is where the quark-gluon plasma is prevalent.  The green area indicates 
the cross over to the hadronic phase that is driven by the presence of excited baryons. Electron 
accelerators CEBAF, ELSA and MAMI have the energy to access this region (color highlights added by author.) }
\label{universe}       
\end{figure}
\section{Introduction}
\label{intro}
\begin{figure}[hb]
\hspace{-0.3cm}\includegraphics[width=9.0cm,height=8.5cm,clip]{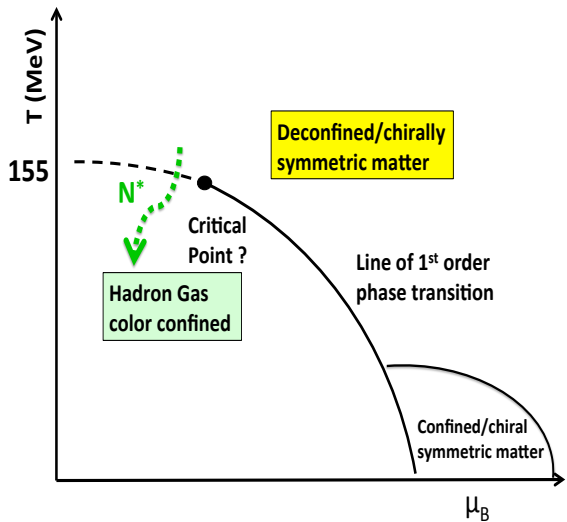}
\vspace{-1cm}\caption{Generic QCD phase diagram showing the de-confined quark-gluon plasma phase (yellow shade) and the 
hadron gas phase (green shade). The thick green line indicates a possible path from the 
de-confined to the confined phase due to the presence of excited baryons bypassing the line of first order phase transition.  }
\label{phase_diagram}      
\end{figure}

Dramatic events occurred in the evolution of the microsecond old universe that had tremendous implication for 
the further development of the universe to the state it is in today. As the universe expanded and cooled sufficiently
into the GeV range (see Fig.~\ref{universe}), the transition occurred from the phase of free (unconfined) quarks and 
gluons, to the hadron phase with quarks and gluons confined in volumes of $\approx 1$~fm$^3$, i.e. protons, neutrons, and 
other baryons. In course of this process, elementary, nearly massless quarks acquire dynamical mass due to 
the coupling to the dressed gluons, and the fact that chiral symmetry is broken dynamically~\cite{Cloet:2013jya}. 
This transition is not a simple first order phase transition, but a "cross over"  between two phases 
of strongly interacting matter, which is moderated by the excitation of baryon resonances starting 
from the highest mass states and ending with the low mass resonances. Figure~\ref{phase_diagram} shows a generic QCD 
phase diagram indicating the region of the cross over from the de-confined region to the confined region of hadrons. 

A quantitative understanding of this transition requires more excited baryons
of all flavors, than have currently been included in the "Review of Particle Properties (RPP)". The full set of states predicted by
quark models with $SU(6)$ symmetry and by Lattice QCD are needed~\cite{Bazavov:2014xya,Bazavov:2014yba} to describe   
the evolution process. The presence of the full complement 
of excited baryons, the acquisition of dynamical mass by light quarks, and the transition from unconfined quarks to 
confinement are intricately related and are at the core of the problems we are trying to solve in hadron physics today. 
We do have all the tools at our disposal to study the individual excited states in relative isolation, and to probe the quark 
mass versus the momentum or distance scale, and to search for so far undiscovered baryon states. 

Accounting for the excitation spectrum of protons and nucleons and accessing many facets of strong interaction mechanisms 
underlying the generation of excited baryons and their intrinsic structures, is one of the 
most important and certainly the most challenging task in hadron physics. It has been the focus of the CLAS N* 
program at Jefferson Lab and likely will remain so with its extensions towards higher energies with CLAS12. 
 
\section{Establishing the light quark baryon spectrum}
\label{baryon_spectrum}
Obtaining an accurate account of the full nucleon resonance spectrum is the basis for making progress in our understanding 
of strong QCD as it relates to light quark sector. We may learn from atomic spectroscopy of the hydrogen atom whose series
of sharp energy levels could be explained within the Bohr model. However, deviations from this model eventually led to the 
QED, which is applicable to all atoms. In the same vein, we need precise measurements of the nucleon excitation spectrum 
to test our best models. The symmetric quark model provides a description of the lower mass spectrum in terms of isospin
and spin-parity quantum numbers, but masses are off, and there are many states predicted within the $SU(6)\otimes O(3)$
symmetry that are missing from the observed spectrum.  Although we have already the correct theory, we cannot really test 
it on the nucleon spectrum, because the full spectrum is not known, and the theory is currently not in a position to predict 
more than what the quark model already has done. Our task is therefore two-fold: 1) to establish the experimental 
nucleon spectrum, and 2) to develop strong QCD to be able to reliably explain it in detail, including masses and hadronic and 
electromagnetic couplings.       
\begin{figure}[t]

\includegraphics[width=8.5cm,clip]{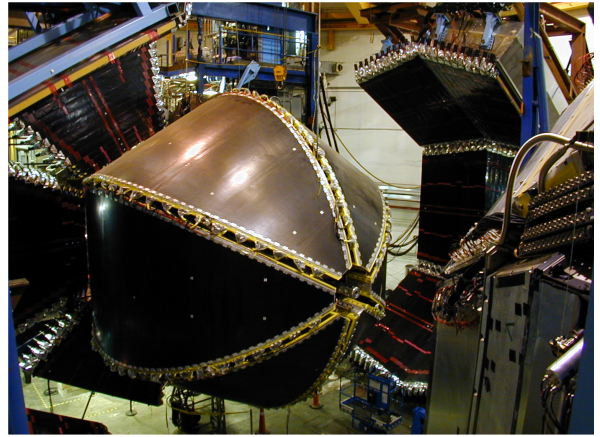}
\caption{The CLAS detector in an open maintenance position exposing the large drift chambers covering nearly 
the full polar angle range. The time-of-flight scintillator bars and the forward electromagnetic calorimeters are also
visible.}
\label{clas}      
\end{figure}
The experimental part has been the goal of the $N^*$ program with CLAS detector (see Fig.~\ref{clas}) and with other facilities, 
especially CB-ELSA. Significant progress has been made in recent years that also  included development of 
multi-channel partial wave analysis frameworks. Much of recent progress came as a result of precise data, including measurement 
of polarization observables collected in the strangeness channel, which I will discuss in the following section.

\begin{figure}[ht]

\hspace{-1cm}
\includegraphics[width=10.0cm,height=10.0cm,clip]{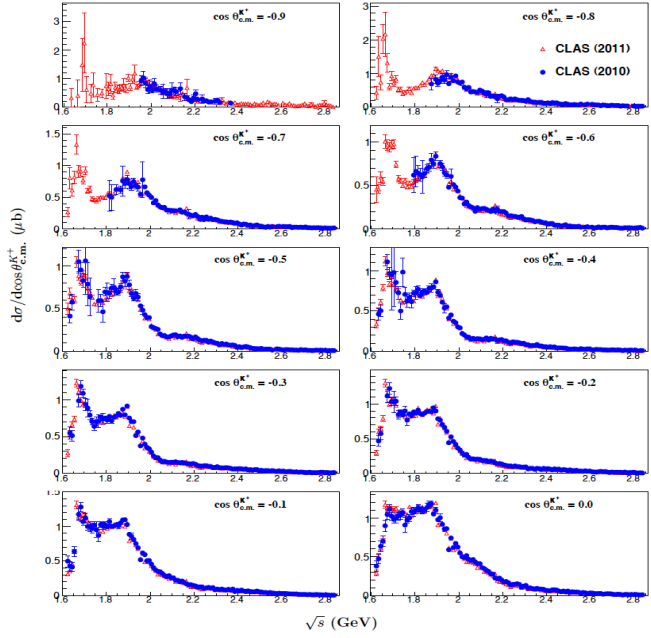}
\vspace{-1.0cm}\caption{CLAS cross section data on $\gamma p \to K^+\Lambda$ differential cross section in the 
backward polar angle range. There are 3 structure visible that indicate resonance excitations, 
at 1.7, 1.9, and 2.2 GeV. The blue full circles are based on the topology $K^+p\pi^-$, 
the red open triangles
are based on topology $K^+p$ or $K^+\pi^-$, which extended coverage towards lower W at backward angles
and allows better access to the resonant structure near threshold. }
\label{KLambda-crs}
\end{figure}

\subsection{Hyperon photoproduction}
\label{hyperon}
 
\begin{figure}[ht]

\hspace{-0.5cm}\includegraphics[width=9.0cm,height=8.5cm,clip]{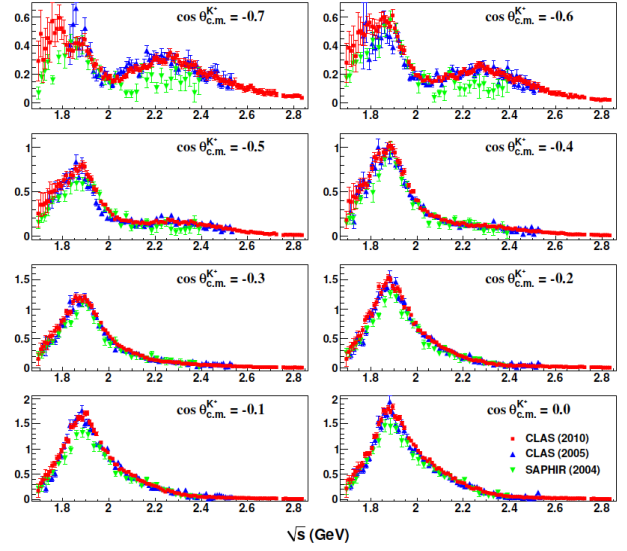}
\vspace{-1.0cm}\caption{Invariant mass dependence of the $\gamma p \to K^+\Sigma^\circ$ differential cross section in 
the backward polar angle range. At the most backward angles there are significant discrepancies between the older 
SAPHIR data and both CLAS data sets, in particular in the mass range from 2.1 to 2.4 GeV, and at backward angles. }
\label{KSigma-crs}
\end{figure}

Here one focus has recently been on precision measurements of the $\gamma p \to K^+\Lambda$ and 
$\gamma p \to K^+\Sigma^\circ$ differential cross section~\cite{Bradford:2006ba,McCracken:2009ra}; 
and using polarized photon beams, with circular or linear 
polarization~\cite{McNabb:2003nf,Bradford:2005pt,McCracken:2009ra,Bradford:2006ba,Dey:2010hh}, 
several polarization observables can be 
 measured by analyzing the weak decay of the recoil $\Lambda \to p \pi^-$, and $\Sigma^\circ \to \gamma \Lambda$. 
 Samples of the data are shown in Fig.~\ref{KLambda-crs} and Fig.~\ref{KSigma-crs}. 
 It is well known  that the energy-dependence of 
 a partial-wave amplitude for one particular channel is influenced by other reaction 
channels due to unitarity constraints. To fully describe the energy-dependence 
of an amplitude one has to include other reaction channels in a coupled-channel approach. 
Such analyses have been developed by the Bonn-Gatchina group~\cite{Anisovich:2011fc}, 
at EBAC~\cite{JuliaDiaz:2007fa}, by the Argonne-Osaka group~\cite{Kamano:2013iva}, 
and the J\"ulich/GWU group~\cite{Ronchen:2014cna}.   

\begin{figure}[ht]
\hspace{-0.5cm}\includegraphics[width=9.0cm,height=8.5cm,clip]{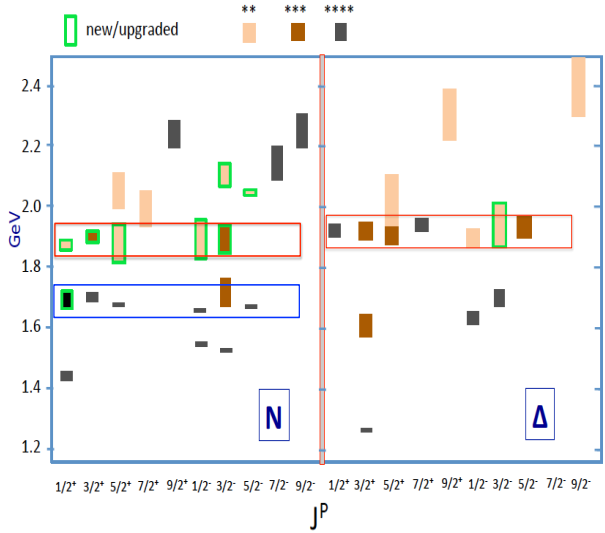}
\vspace{-1.0cm}\caption{The nucleon and $\Delta$ spectrum for masses up to 2.4~GeV. The green frames indicate 
states that have newly been found in the BnGa analysis or have been upgraded in their star rating in the 
RPP~\cite{Agashe:2014kda}. The red and blue boxes highlight near mass degenerate
states with different spin and parity. The states at 1.9 GeV are all from the latest additions to RPP. They seem to repeat a 
pattern of states observed near 1.7 GeV. A similar pattern is observed in the $\Delta$ sector at 1.9 GeV, with a new addition at $J^P = {3\over 2}^-$ filling in a void of  in the previously observed pattern. }
\label{BaryonStates2016}
\end{figure}

The Bonn-Gatchina group has claimed a set of eight states that are either newly discovered or have significantly improved 
evidence for their existence.  These states entered in recent editions of the Review of Particle Properties~\cite{Agashe:2014kda}. Figure~\ref{BaryonStates2016} shows the nucleon and $\Delta$ spectrum with the new additions highlighted.

\begin{figure}[hb]
\hspace{-0.2cm}\includegraphics[width=8.5cm,height=8cm,clip]{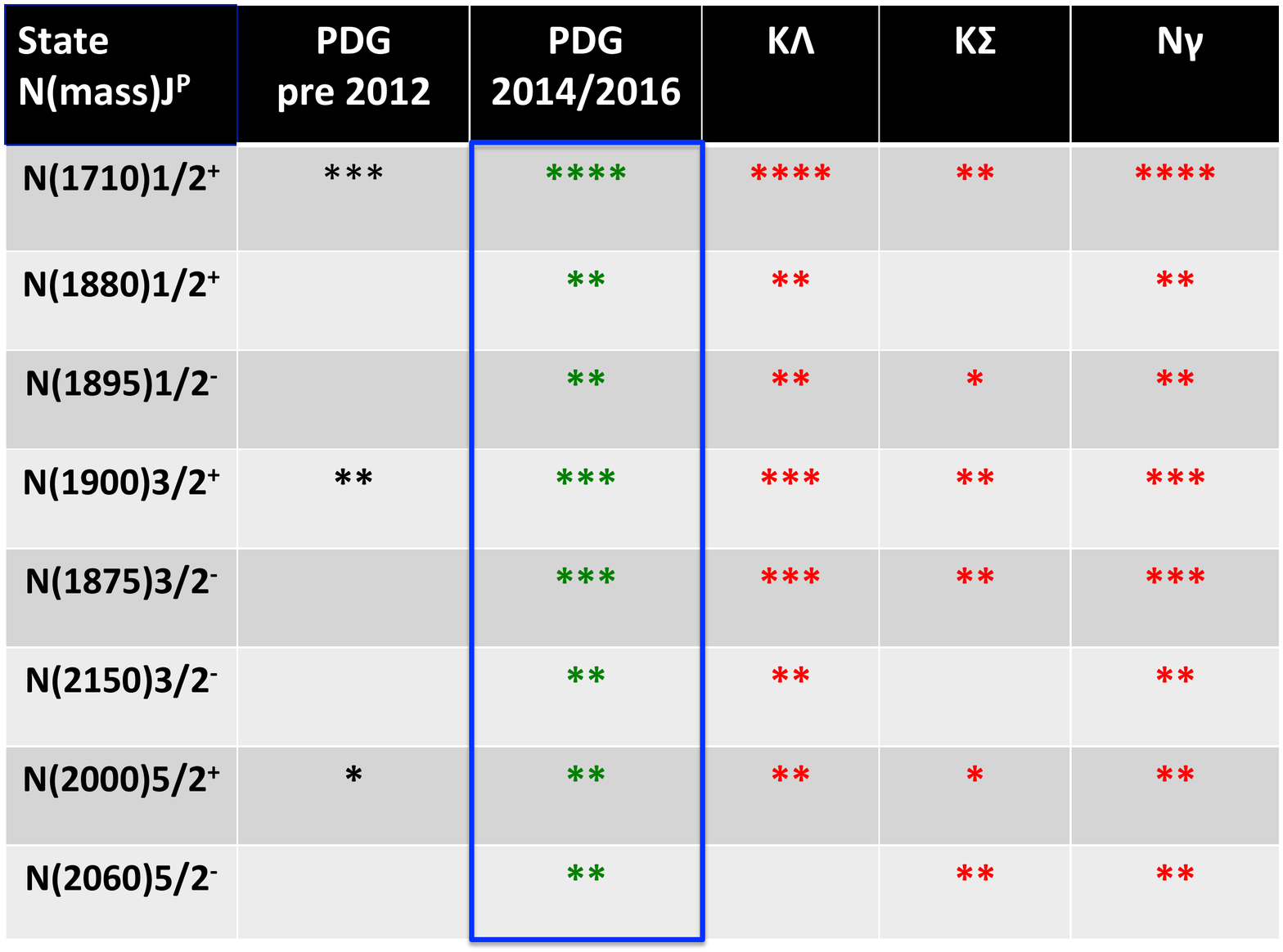}
\caption{Recently discovered nucleon resonances with their star ratings in the Review of Particle Properties, 2014/16. 
The states have been observed in the Bonn-Gatchina multi-channel partial wave analysis of photo-produced 
$K^+\Lambda$ and $K^+\Sigma^\circ$ final states.}
\label{pdg2016}      
\end{figure}

Figure~\ref{pdg2016} shows the nucleon resonances observed in the Bonn-Gatchina multi-channel partial wave analysis
using the hyperon photoproduction data from CLAS and other data sets. Figure~\ref{BaryonStates2016} shows
 how the new states fit into the previously observed nucleon and $\Delta$ states for 
masses up to 2.4~GeV.  

\begin{figure}[h]

\hspace{-0.5cm}\includegraphics[width=9.0cm,height=8.5cm,clip]{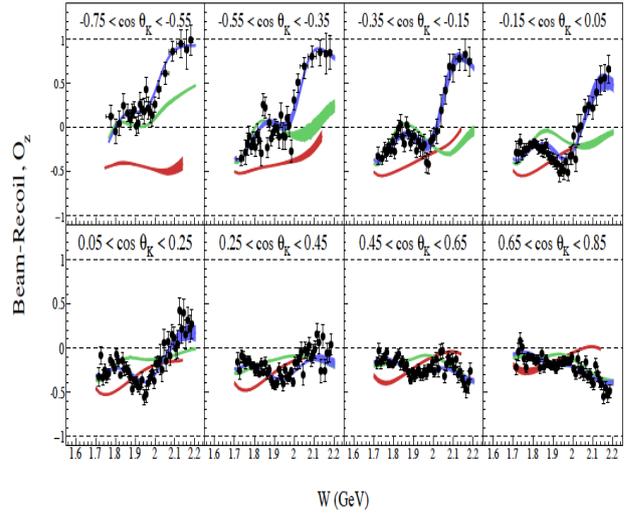}
\vspace{-1.0cm}\caption{Double polarization observable $O_z$ for different $K^+$ polar angle bins. The bands are 
projections  by ANL-Osaka (red), BnGa 2014 (green), and BnGa 2016 (blue). The latter included the new data in the fit. 
The large discrepancies at $W > 2$GeV indicate possible high mass resonance strength that was not included in earlier fits. }
\label{KLambda_Oz}
\end{figure}

New data on $K^+\Lambda$ production using a linearly polarized photon beam and measuring the $\Lambda$ 
recoil polarization along the $\Lambda$ momentum~\cite{Paterson:2016vmc} are shown in Fig~\ref{KLambda_Oz}. 
These data show strong sensitivity to excited baryon states, including possible new states, but they have 
not been included in previous full multi-channel partial wave analyses. 

\subsection{Vector meson photoproduction}
\label{vectormesons}
A large volume of precision photoproduction data have been taken with CLAS on $\gamma p \to p \omega$ 
covering the mass range from threshold to 2.8 GeV~\cite{Williams:2009ab}. The final state has been 
measured in its dominant decay channel $\omega \to \pi^+ \pi^- \pi^\circ$.  
Using neutral-charge vector mesons in partial wave analyses is more involved 
compared to pseudoscalar mesons as the spin $J=1$ meson coupling with the spin 1/2 proton results in 
more amplitudes defining the process. The advantage, however, is that the process can only proceed through isopsin-1/2
nucleon resonances, which simplifies the analysis.  There are also large t-channel and diffractive processes that 
give large background contributions. As of this writing this data set has not yet been included in a multi-channel partial 
wave analysis. 

\begin{figure}[ht]
\hspace{-0.5cm}
\vspace{-1.0cm}\includegraphics[width=9.5cm,height=8.0cm,clip]{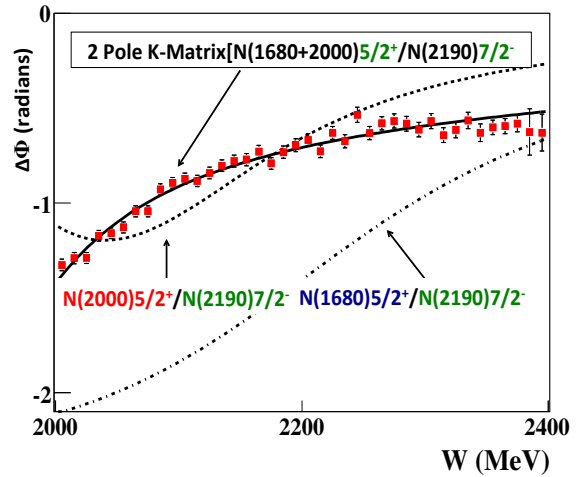}
\caption{The phase motion of $\gamma p \to p \omega$ with the $\omega$ fully reconstructed from its decay 
$\omega \to \pi^+ \pi^- \pi^\circ$ and the differential cross section and (unpolarized) spin-density matrix elements 
determined from the data.}
\label{omega}      
\end{figure}

In spite of these complications, the omega data have been employed in a single channel event-based analysis 
and fit in a 2-pole K-matrix procedure~\cite{Williams:2009aa}. Figure~\ref{omega} demonstrates the sensitivity
of this channel to $N(2000){5\over 2}^+$, a 2-star candidate state whose existence is strongly supported by this analysis. 

\begin{figure}[ht]
\hspace{-1.3cm}
\vspace{-0.5cm}\includegraphics[width=11.5cm,height=10.0cm,clip]{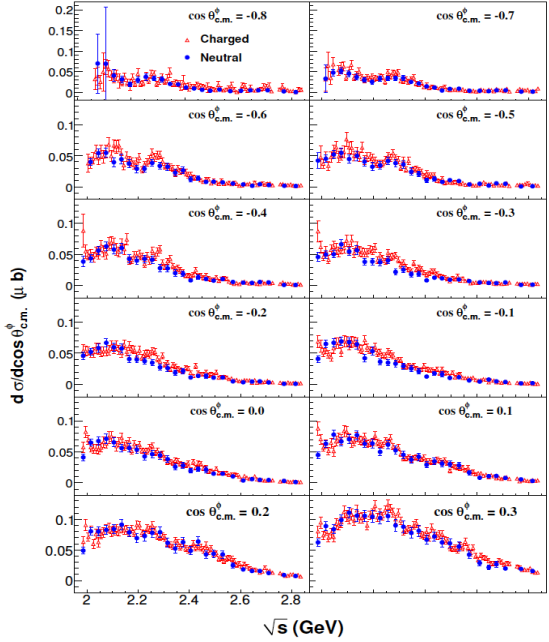}
\caption{Differential cross section of $\phi$ photoproduction, shown in the backward hemisphere. The open red circles 
are from the charged final state, the solid blue points from the neutral charge final state. There are enhancements
in both charge channels at the most backward angle bins near $\sqrt{s}= 2.25$~GeV. }
\label{phi_crs}      
\end{figure}

The process $\gamma p \to p \phi$ has been measured in its dominant final state  $K \bar{K}$ with both charge channels 
 $\phi \to K_s^\circ K_l^\circ$~\cite{Seraydaryan:2013ija}, and $\phi \to K^+ K^-$~\cite{Dey:2014tfa}. Figure~\ref{phi_crs} 
 shows the cross section in the backward hemisphere. There are small enhancements near $\sqrt{s}=2.25$~GeV, indicating resonance structure that is however overwhelmed by the increasing background contributions at more forward angles.
\begin{figure}[ht]
\hspace{-0.7cm}
\vspace{-0.5cm}\includegraphics[width=9.0cm,height=8.0cm,clip]{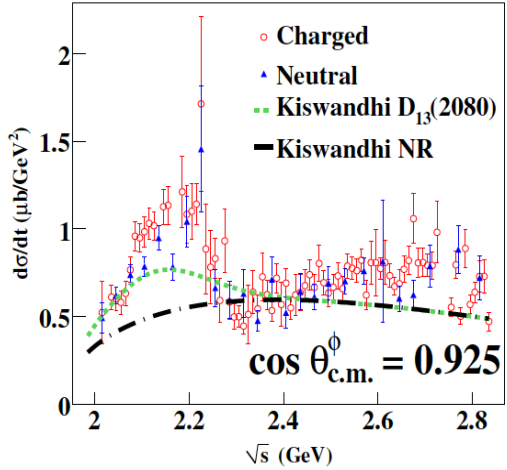}
\caption{Differential cross section $\gamma p \to \phi p$ in the forward most angle bin. }
\label{phi_forward_peak}      
\end{figure}
 
 There is a strong bump near $\sqrt{s}=2.2$~GeV seen in Fig.~\ref{phi_forward_peak} at the forward most angle bin. 
 The bump is present in both charge channels and does not appear to be due to a genuine baryon resonance as it 
 is not present at larger angles and does 
 not have a partial wave content that would be typical for a baryon resonance with specific spin and parity. There are 
 model speculations~\cite{Lebed:2015fpa,Lebed:2015dca} that the structure might be due to an diquark-antitriquark 
 in a $(su)(\bar{s}ud)$ configuration, which could explain the strong forward bump correlated with much smaller 
 enhancements at backward angles. 
 A reaction model and partial wave analysis will be needed to come to a more definitive interpretation of the observed 
 cross section behavior.  

\section{Electroexcitation of light-quark baryon resonances}
As high precision photoproduction processes have been successfully used to search for new baryon states, and will continue 
in searches for new states at higher masses, we also need to exploit electron scattering to probe the interior of 
excited states and understand their intrinsic spatial and spin structure. 
Meson electroproduction has revealed intriguing new information regarding the active degrees of freedom underlying 
the structure of the excited states and their scale dependences. At short distances the 
resonance electrocouplings are sensitive to the momentum dependence of the light quark masses as predicted in DSE
calculations starting from the QCD Lagrangian~\cite{Segovia:2014aza}. They have also been incorporated in the $Q^2$ dependence of the quark masses in the LF RQM~\cite{Aznauryan:2012ec}. At larger distances the role of meson-baryon 
contributions and of meson cloud effects become relevant.  
\begin{figure}[ht]
\centering

\includegraphics[width=8.0cm,height=6.0cm,clip]{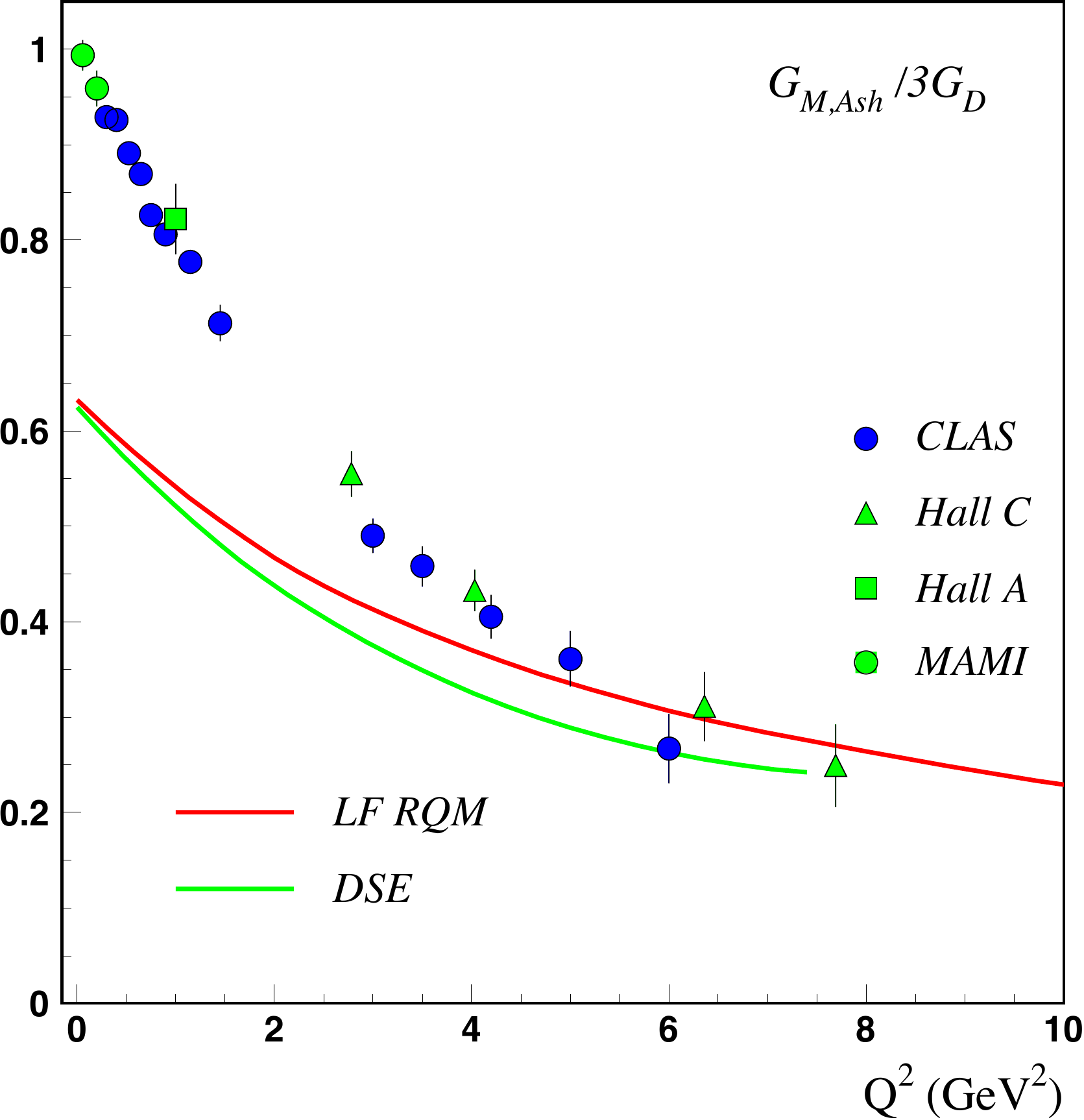}
\includegraphics[width=8.0cm,height=8.5cm,clip]{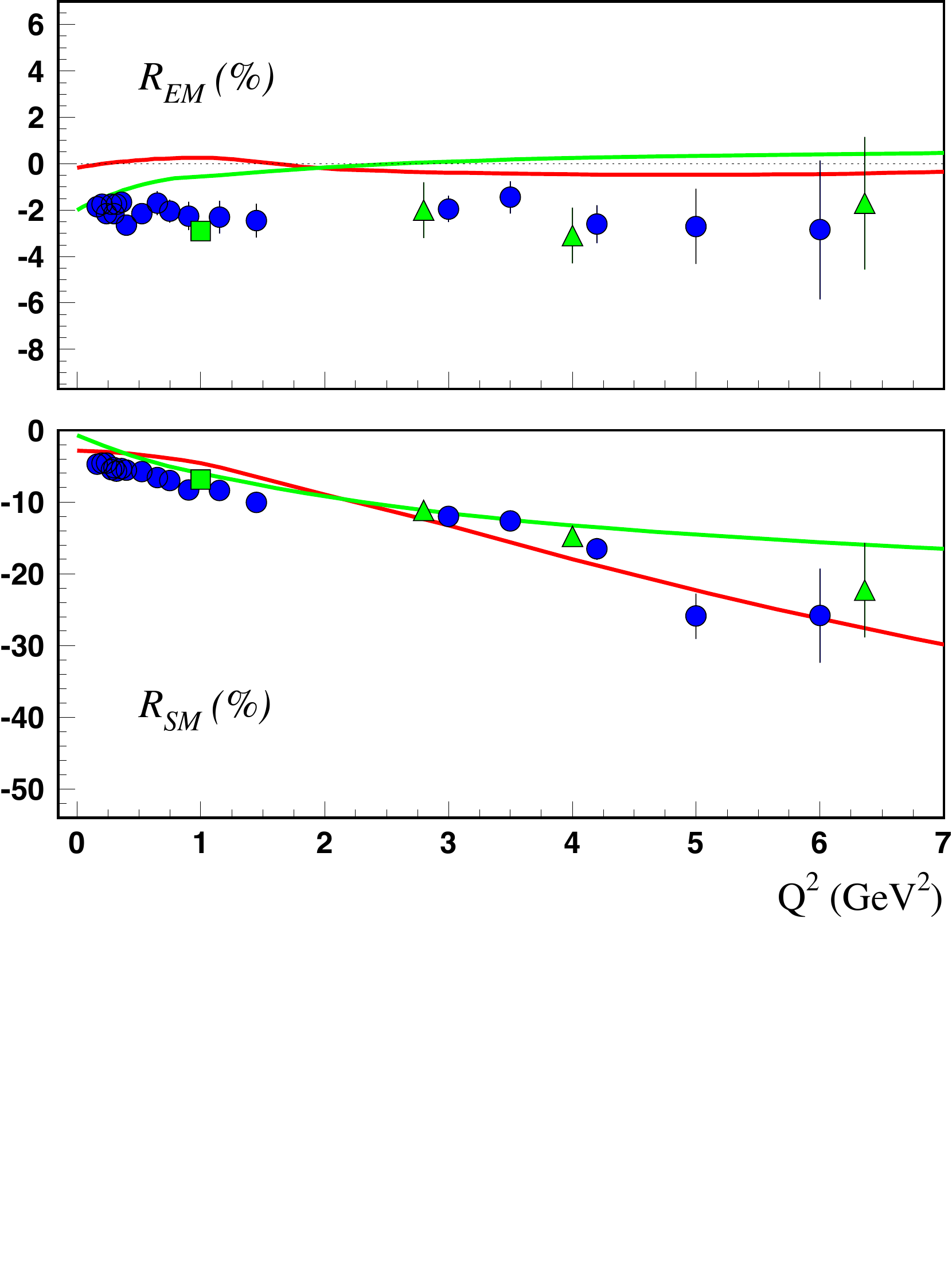}
\vspace{-2.4cm}\caption{\small The $N\Delta$ magnetic transition form factor (top) and the electric and scalar 
quadrupole ratios $R_{EM}$ and $R_{SM}$ from CLAS (blue circles), Hall C (triangles) and Hall A (square) 
experiments from $ep \to ep\pi^0$. The green circles are from photoproduction data. 
The green line is the result of the first principles DSE calculation~\cite{Segovia:2014aza}. The red line is the prediction of the
  LF RQM~\cite{Aznauryan:2015zta} with running 
quark mass and configuration mixing~\cite{Aznauryan:2016wwm}. The model is normalized to the data at 
$Q^2 > 4$~GeV$^2$ where the 3-quark core contributions dominate. }
\label{fig:Delta}
\end{figure}
\subsection{The $\rm N\Delta(1232)$ transition}

One of the important insights emerging from extensive experimental and theoretical research 
is clear evidence that resonances are not excited 
from quark transitions alone, but there can be significant contributions from 
meson-baryon interactions as well, and that these two processes contribute to the 
excitation of the same state. This evidence has been obtained in part through 
the observation that the 
quark transition processes  often do not have sufficient strength to explain fully 
the measured resonance transition amplitudes at the real photon point. The best studied 
example is the 
$\Delta(1232){3\over 2}^+$ resonance, which, when excited electromagnetically, is dominantly 
due to a magnetic dipole transition from the nucleon ground state, but only about 2/3  
of the transverse helicity amplitudes at the photon point can be explained by the $q^3$ content of the state.
In contrast, at $Q^2 \ge 3$~GeV$^2$ the quark contribution is nearly exhausting the measured strength as 
can be seen in Fig.~\ref{fig:Delta}.

At low $Q^2$ a satisfactory description of this transition has been achieved in hadronic models that include 
pion-cloud contributions  
and also in dynamical reaction models, where the missing strength has been attributed 
to dynamical meson-baryon interaction in the final state. 
A recent calculation within the light front relativistic quark model (LF RQM)~\cite{Aznauryan:2016wwm} 
with the $q^3$ contributions normalized to the high $Q^2$ behavior, finds the meson-cloud contributions 
to set in at $Q^2 \leq 3.0$~GeV$^2$ as shown in Fig.~\ref{fig:Delta}.    
The excitation of this and other states using electron scattering should be highly sensitive to the different Fock 
components in the wave function of these excited states as it is expected that they have different
excitation strengths when probed at large and at short distance scales, 
i.e. with virtual photons at low and high $Q^2$. At high $Q^2$ we expect 
the $q^3$ components to be the only surviving contributions, while the higher Fock states may have large, 
even dominant strength at low $Q^2$.     
\begin{figure}[t]
\hspace{0.5cm}\includegraphics[width=7.0cm,height=6.0cm,clip]{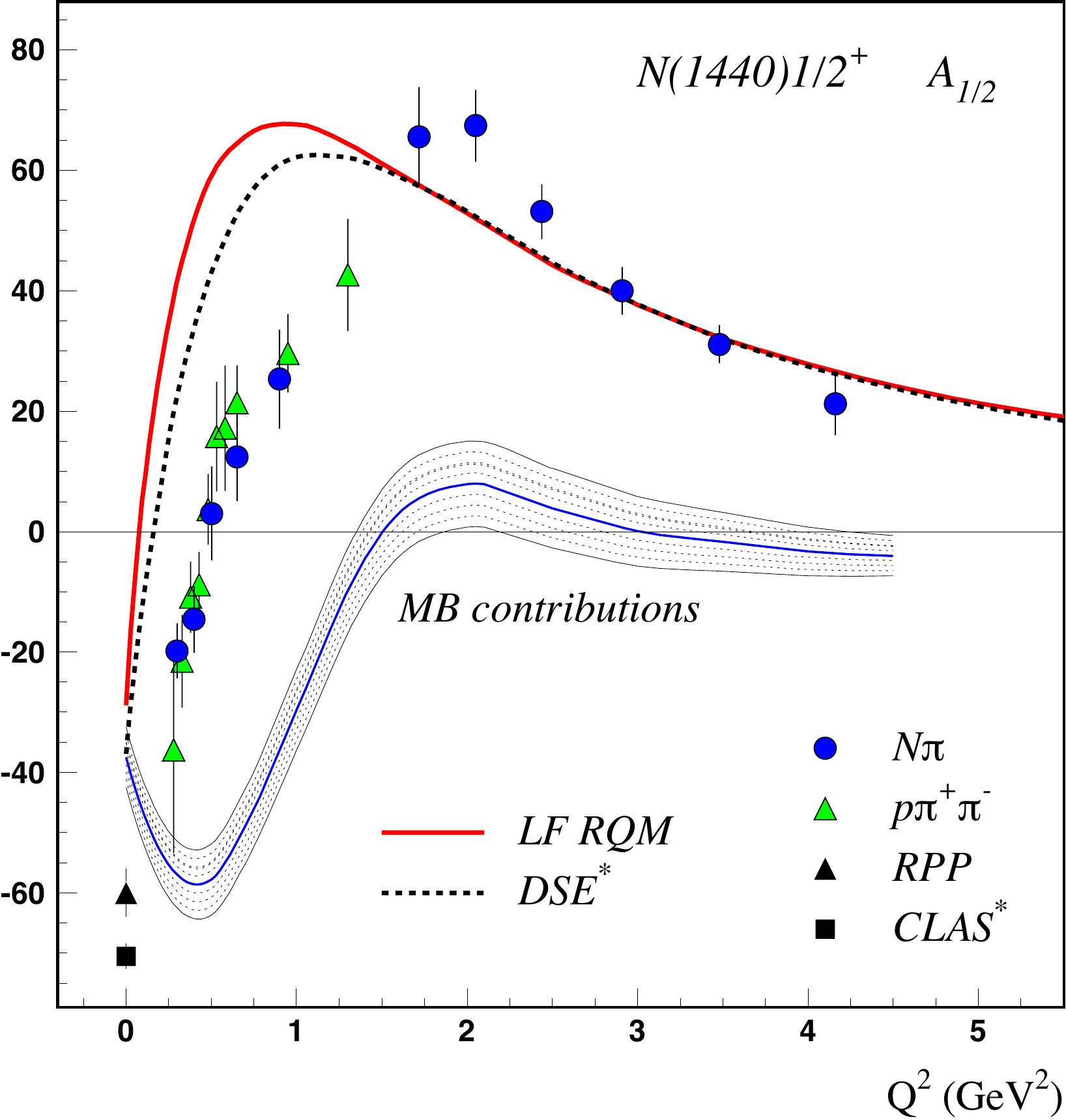}

\hspace{0.5cm}\includegraphics[width=7.0cm,height=6.0cm,clip]{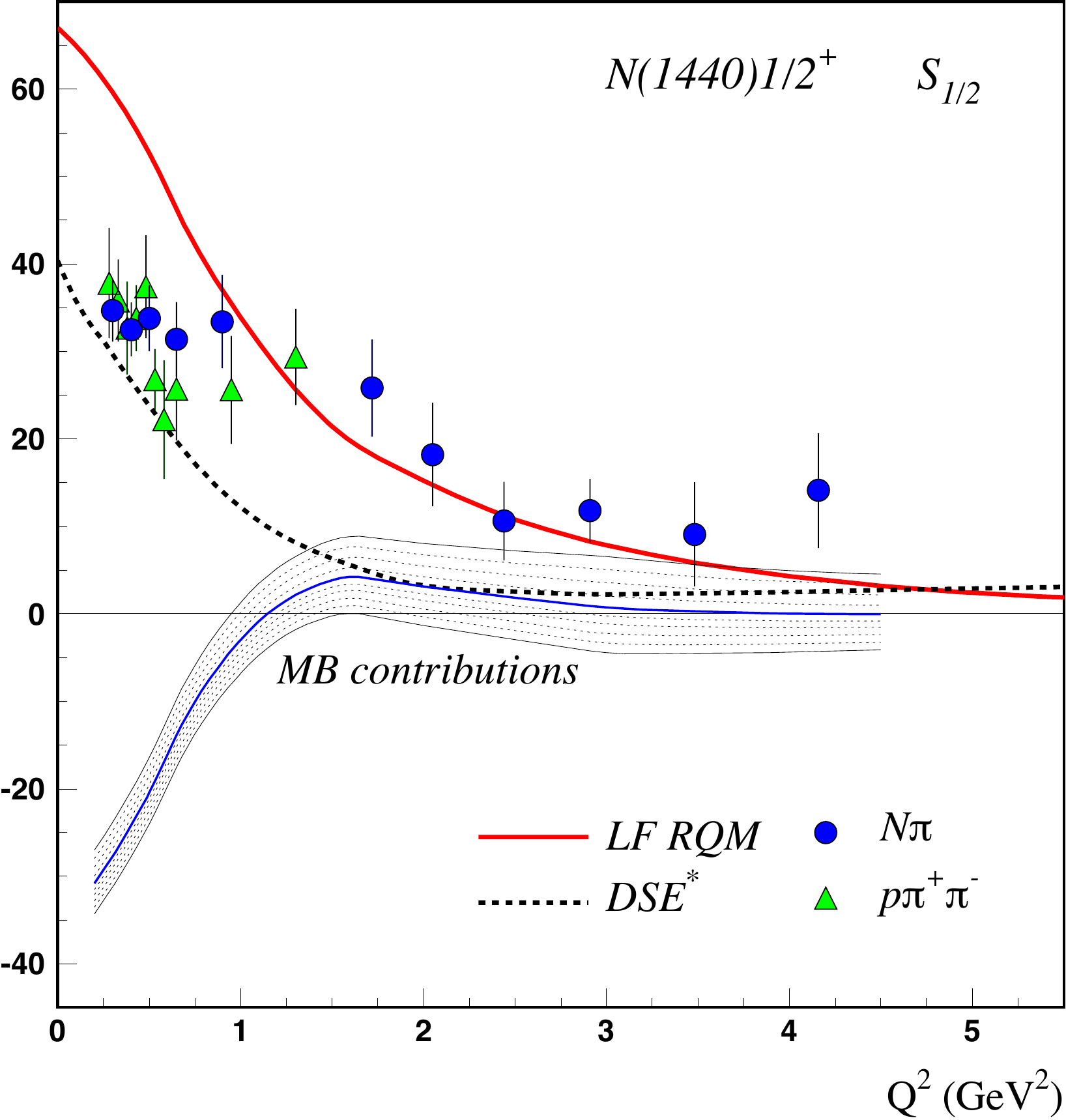}
\caption{\small The electrocoupling amplitudes $A_{1/2}(Q^2)$ and $S_{1/2}(Q^2)$ for the Roper resonance 
$N(1440){1\over 2}^+$ (for units see Fig.~\ref{fig:Delta}) . 
Electroproduction data are from 
CLAS~\cite{Aznauryan:2008pe,Aznauryan:2009mx,Mokeev:2012vsa,Mokeev:2015lda}. The dashed curve is 
from the DSE/QCD 
calculation~\cite{Segovia:2015hra} after renormalization of the quark core contribution at high $Q^2$.
The solid curve is from the LF RQM with momentum-dependent quark mass~\cite{Aznauryan:2012ec,Aznauryan:2016wwm}. 
The shaded band represents the inferred meson-baryon contributions. Although there is some model-dependence
 concerning the precise forms of these contributions, the marked similarity between those inferred via the LF RQM 
 and those determined in a parameter-free DSE analysis~\cite{Roberts:2016dnb} indicates that a quantitative 
 understanding of these effects is near.}
\label{fig:a12roper}
\end{figure}

\subsection{Solving the Roper puzzle}
\label{sec:roper}
It is well known that the Roper $N(1440){1\over 2}^+$ state presented the 
biggest puzzle of the prominent nucleon resonances and for decades defied explanations within models.  
The non-relativistic constituent quark model has it as the first radial excitation of the 
nucleon ground state. However, its physical mass is about 300 MeV lower than what is predicted. 
The most recent LQCD projections have the state even 1 GeV above the nucleon ground state, 
i.e. near 1.95GeV~\cite{Edwards:2011jj}. 
The electromagnetic transition amplitude extracted from pion photo production data is large and negative, while 
the non-relativistc constituent quark model (nrCQM) predicts a large and positive amplitude. 
Furthermore, the early electroproduction results showed a rapid disappearance of its  
excitation strength at $Q^2 \leq 0.5$GeV$^2$, while the model predicted a strong rise in 
magnitude. These apparent discrepancies led to attempts at alternate interpretations of the state, 
e.g. as the lowest gluonic excitation of the nucleon~\cite{Barnes:1982fj} 
and as dominantly $N\rho$~\cite{Cano:1998wz} or $N\sigma$~\cite{Obukhovsky:2011sc} molecules.    

Recent development of the dynamically coupled channel model by the EBAC group, has  
led to a possible resolution of the discrepancy in the mass value, by including resonance couplings to inelastic 
decay channels in their calculations~\cite{Suzuki:2009nj}. The inelastic channels cause the dressed Roper pole to move 
by over 350 MeV from its bare value of 1.736~GeV to 1.365~GeV, i.e. close to where it is found experimentally. 

\begin{figure}[t]
\hspace{0.5cm}\includegraphics[height=6.0cm,width=7.0cm,clip]{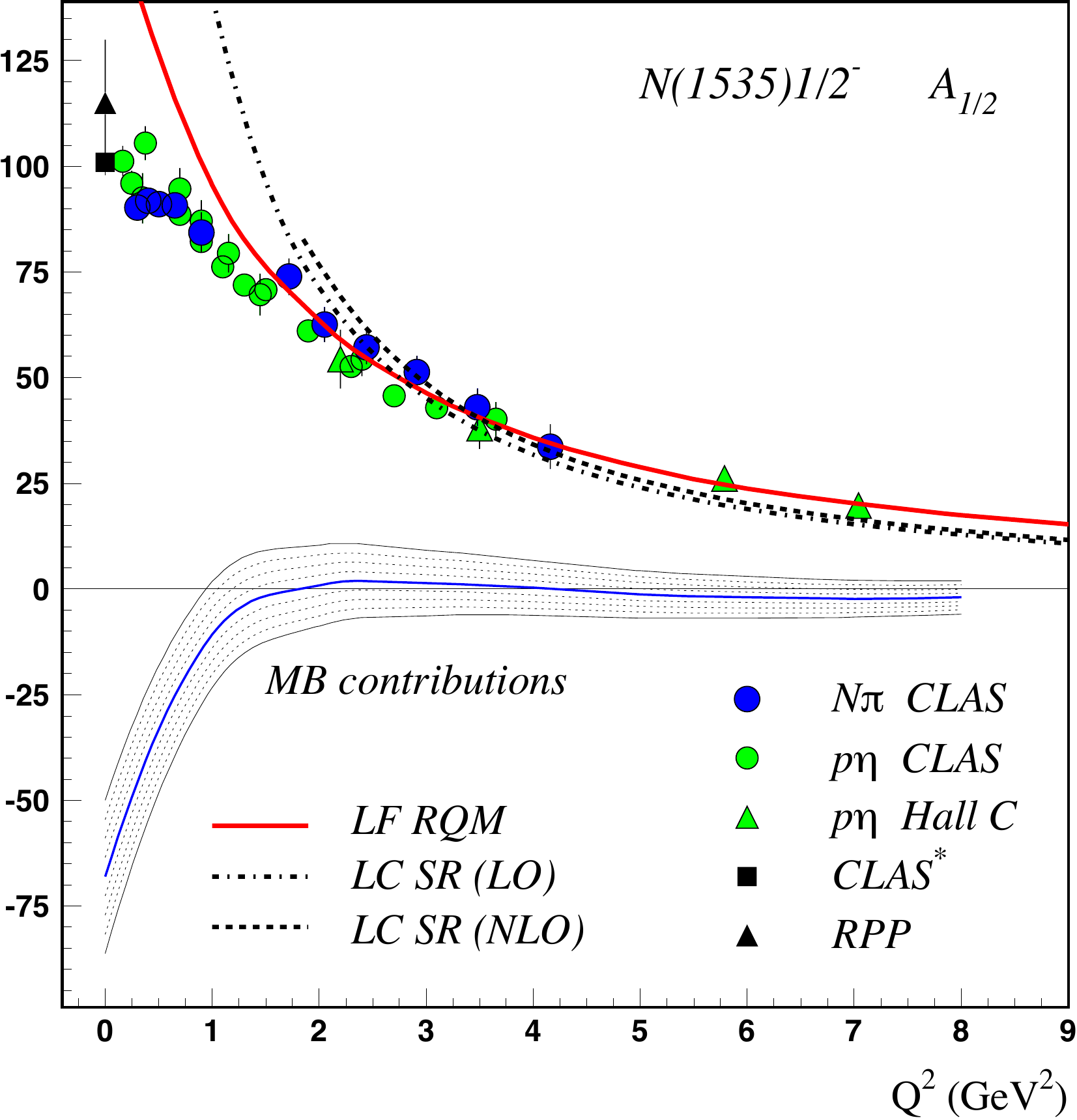}

\hspace{0.5cm}\includegraphics[height=6.0cm,width=7.0cm,clip]{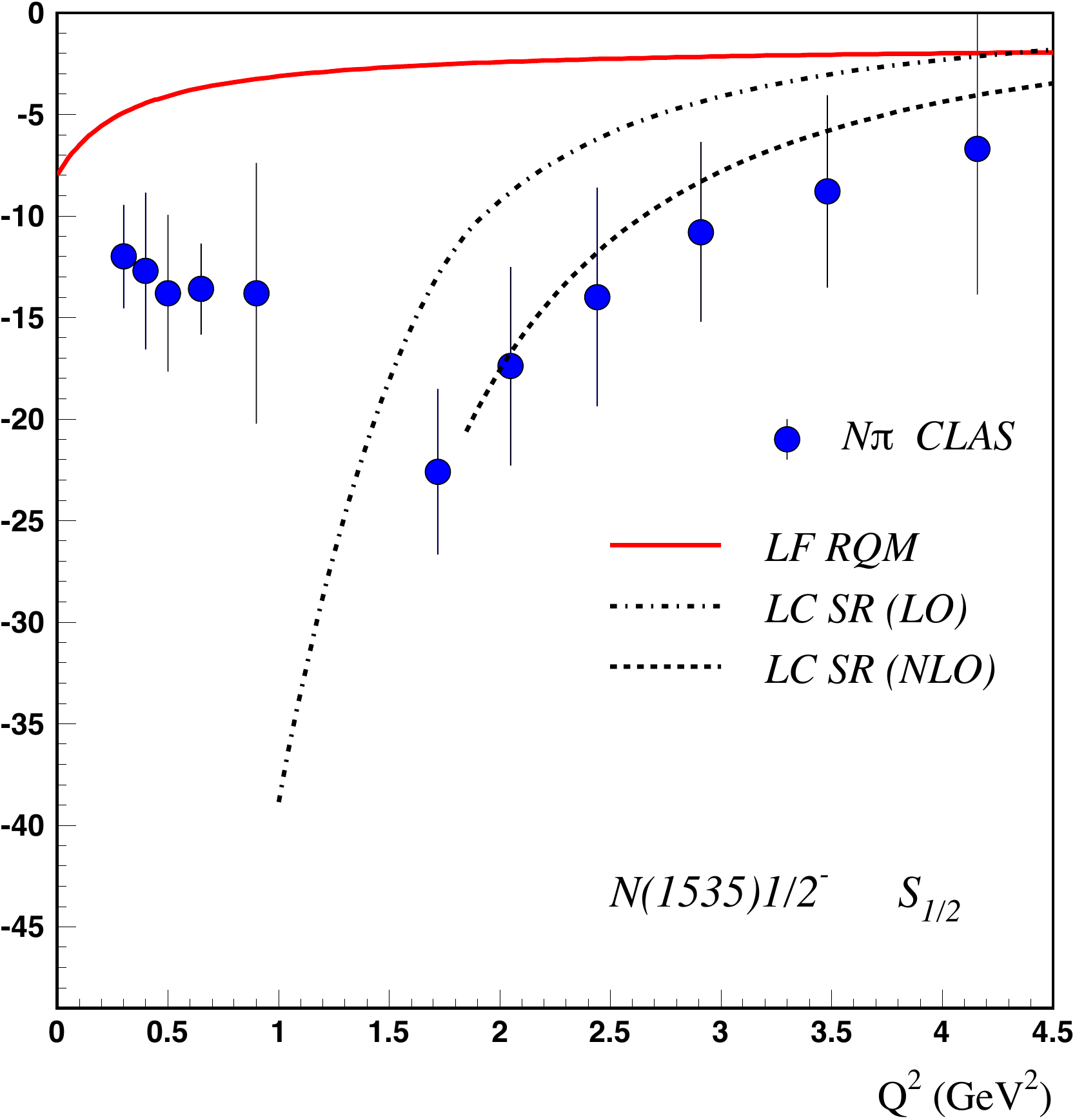}
\caption{\small The transverse and scalar amplitudes for the $N(1535){1\over 2}^-$  determined in 
$ep \to ep\eta$ (open symbols) and in $ep \to eN\pi$ (full circles). 
The shaded bands indicates the size of the meson-baryon contributions evaluated by subtracting 
LF RQM projection from smoothened data points. 
Curves represent LF RQM (solid) and LC SR (dashed-dotted). }
\label{fig:N1535}
\end{figure}
Measurements of meson electroproduction data in a large range of $Q^2$~\cite{Egiyan:2006ks,Park:2007tn,Fedotov:2008aa,Aznauryan:2009mx} 
provided the basis for analyses of the electrocoupling amplitudes of the $pN(1440){1\over 2}^+$ transition. The electrocoupling 
amplitudes are shown in Fig.~\ref{fig:a12roper}.  
The LF RQM predicts the correct sign of the transverse amplitude at $Q^2=0$ and a sign change at small $Q^2$. The 
behavior at low $Q^2$ is described well when the $q^3$ component in the wave function is complemented by 
meson-baryon contributions, e.g. $N\rho$~\cite{Cano:1998wz} or  $N\sigma$~\cite{Obukhovsky:2011sc}, 
and also in effective field theories~\cite{Bauer:2014cqa} employing pions, $\rho$ mesons, 
the nucleon and the Roper $N(1440){1\over 2}^+$ as effective degrees of freedom. 
The high-$Q^2$ behavior is well reproduced in the QCD/DSE approach and the LF RQM which include 
momentum-dependent quark masses, in QCD/DSE~\cite{Segovia:2015hra} due to the full incorporation 
of the momentum-dependent dressed quark mass in QCD and modeled in LF RQM~\cite{Aznauryan:2012ec}. 
The latter is parameterized as $Q^2$-dependent mass in the quark wave function.   
\subsection{$N(1535){1\over 2}^-$, parity partner of the nucleon}
\label{N1535}
The parity partner of the ground state nucleon lies 600 MeV above the mass of the nucleon. The shift 
is thought to be due to the breaking of chiral symmetry in the excitation of nucleon resonances. 
Figure~\ref{fig:N1535} shows the $A_{1/2}$ and $S_{1/2}$ amplitudes. 
The former is well 
described by the LF RQM~\cite{Aznauryan:2012ec} and the  LC SR (NLO)~\cite{Anikin:2015ita} 
evaluation for $Q^2 \geq 1.0 $GeV$^2$ and $Q^2 \geq 2$GeV$^2$, respectively. 
The scalar amplitude $S_{1/2}$ 
departs from the LF RQM predictions significantly, it is, however, well described by the LC SR (NLO) 
calculation at $Q^2 \ge 1.5$GeV$^2$.  This result points at a promising approach of relating the resonance 
electrocouplings to calculations from first principles of QCD.  
The state has also been discussed as having large strangeness components~\cite{Zou:2006tw}, 
an assertion that might account for the discrepancy in the scalar amplitude with the data at low $Q^2$, 
although no specific predictions for the $S_{1/2}$ amplitude are available that include such contributions.  

\subsection{Light-cone charge transition densities}
\begin{figure}[b]
\hspace{-1.2cm}\includegraphics[height=10.0cm,width=11.0cm,clip]{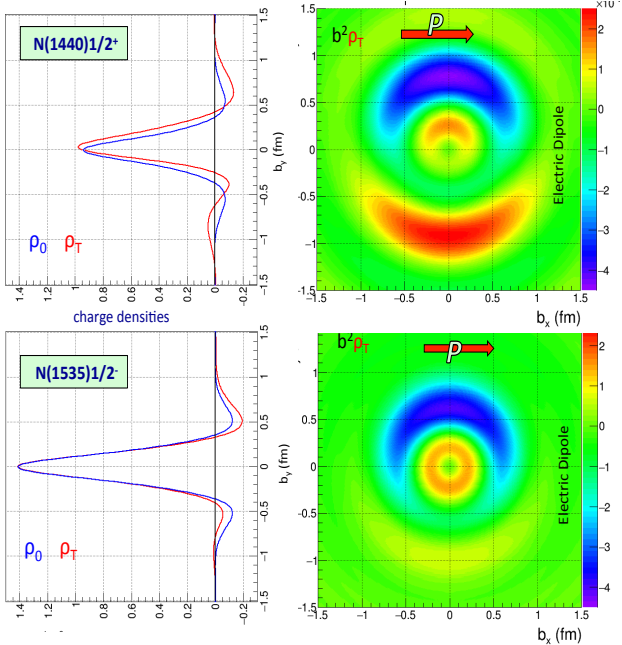}
\vspace{-1.0cm}\caption{\small The light-cone charge transition densities for the Roper resonance (top panels)
 and the $N(1535)$ state. The left panels show projections on the transverse impact parameter for 
 unpolarized protons (blue lines) and for protons polarized along the x-axis (red line). For better visibility the densities have
 been multiplied with $b^2$, which explains the zero in the center $b_{x,y}=0$ and the enhancements at 
 large $b_{x,y}$.  }
\label{charge_densities}
\end{figure}

The electrocoupling amplitudes of the $N(1440){1\over 2}^+$ and $N(1535){1\over 2}^-$ states exhibit quite
different $Q^2$ dependences. To see in what way this reflects the spatial structure of the two states we 
use the prescription in Ref.~\cite{Tiator:2008kd} to perform a Fourier transform of the resonance transition form 
factors by extrapolating the range in $Q^2$, where electrocouplings are known, to infinity. Obviously, this involves some model assumptions, for which we assume  dimensional scaling behavior of the amplitudes. For the $N(1535){1\over 2}^-$ 
we have data up to $Q^2=8$~GeV$^2$, and the extrapolation is relatively safe, less so for $N(1440{1\over 2}^+$ where 
the data end at $Q^2 \approx 5$~GeV$^2$. Asymptotically, the amplitudes of both states must have the same 
power behavior of $A_{1/2} \propto 1/Q^3$. The light-cone transition charge densities are shown in Fig.~\ref{charge_densities}
for the unpolarized transition ($\rho_0$) and for the transition from transversely polarized protons ($\rho_T$), respectively.  
The charge density 
$\rho_0$ of the proton-Roper transition shows clearly a softer center and a more extended "cloud". In the polarized
case, the charge 
center of $\rho_T$ moves farther to positive $b_y$ values, generating a larger electric dipole behavior than is the case for 
$N(1535){1\over 2}^-$. The latter shows a stronger central peak for $\rho_T$, that remains fixed at $b_y=0$. The results 
indicate a stronger {\sl hadronic} component for the Roper resonance than for the $N(1535){1\over 2}^-$ that 
appears more {\sl quark~like}.  
Note that for ease of comparison both scales, dimensions and color codes are the same so that the two resonances can be compared directly.     
\subsection{Spin structure of the $\gamma pN(1520){3\over 2}^-$ transition}
\label{N1520}
\begin{figure}[tbh]
\hspace{0.0cm}\includegraphics[height=6.0cm,width=7.0cm,clip]{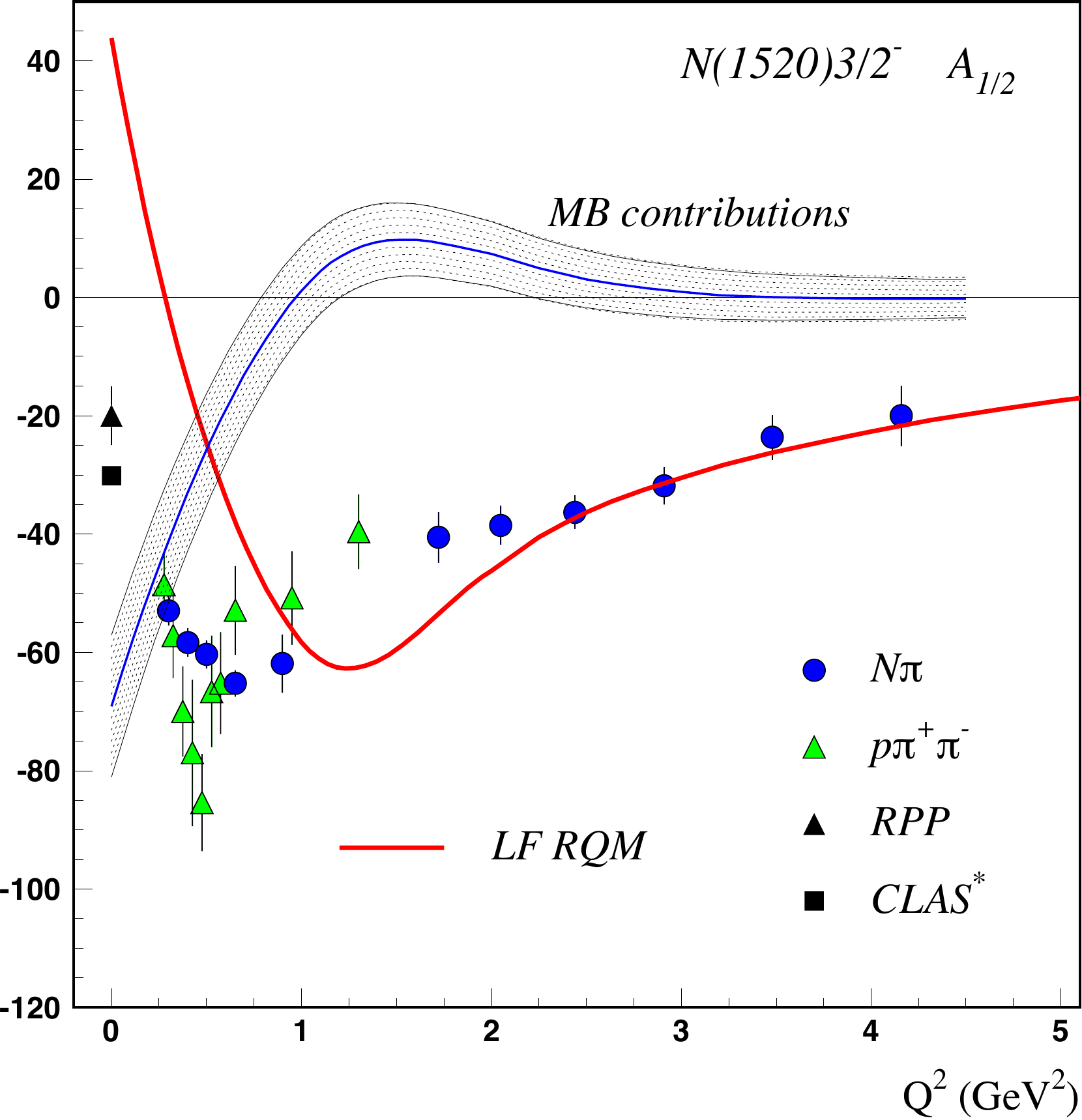}
\hspace{0.5cm}\includegraphics[height=6.0cm,width=7.0cm,clip]{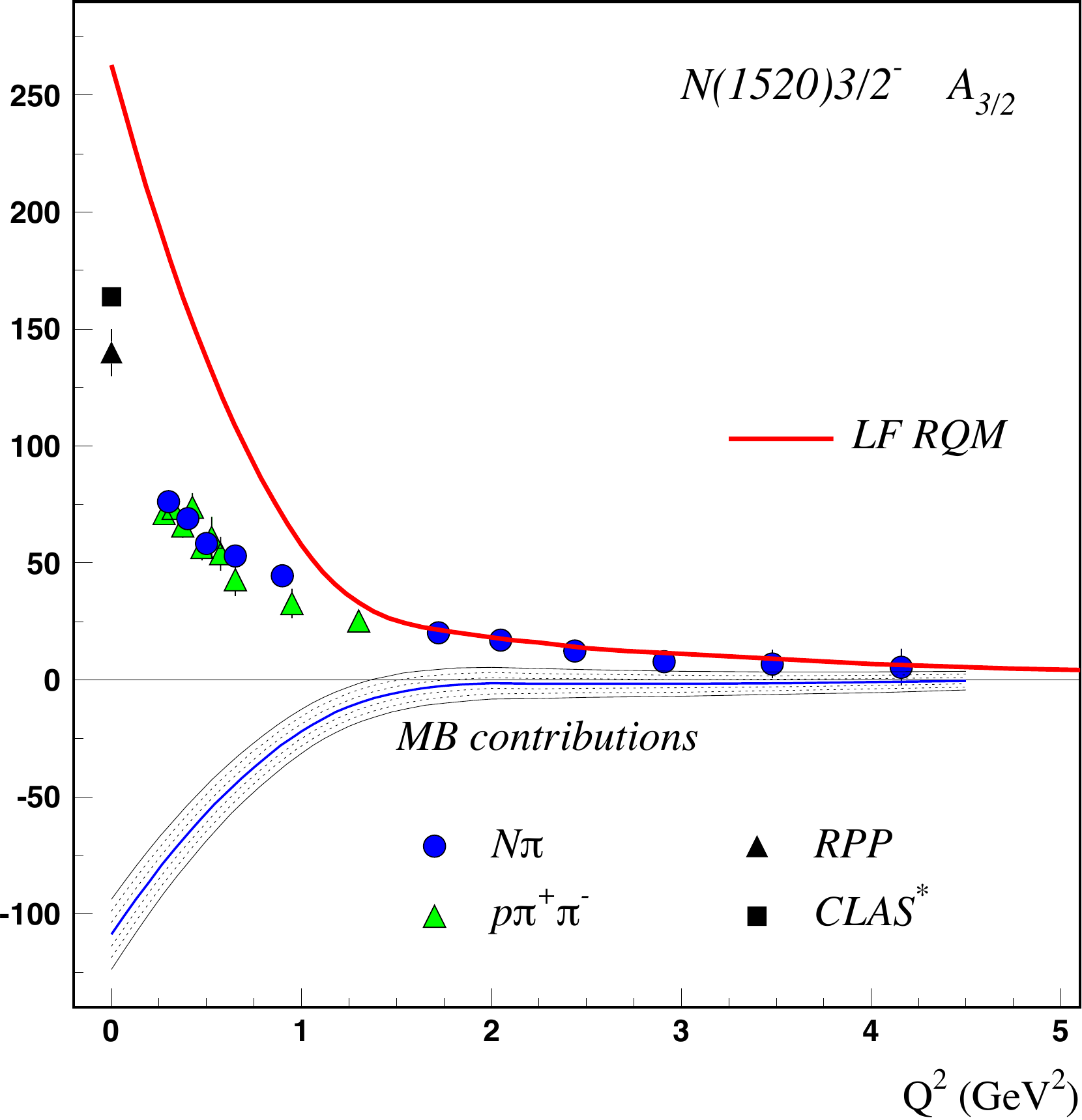}
\hspace{0.5cm}\includegraphics[height=6.0cm,width=7.0cm,clip]{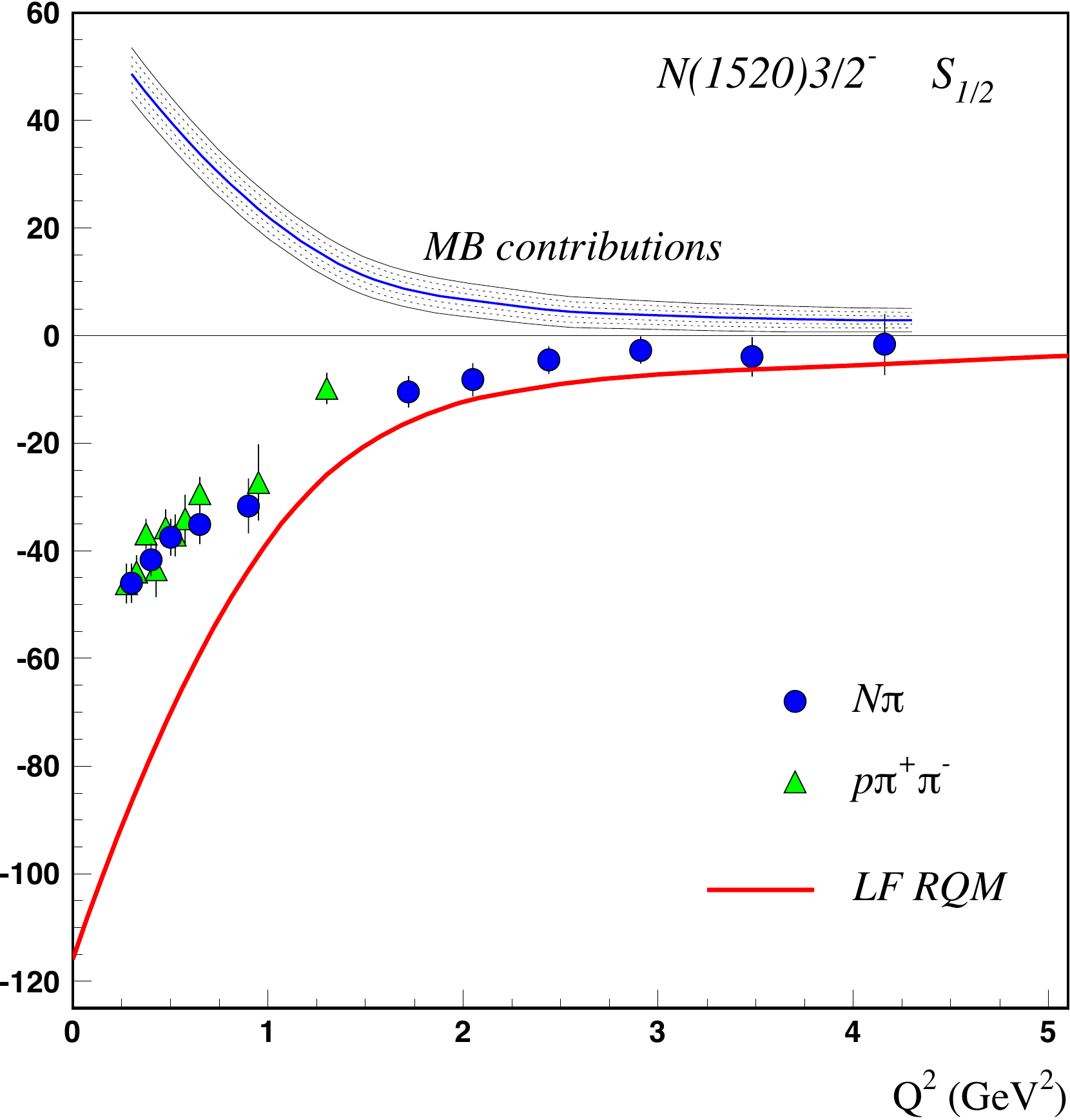}
\caption{\small The transverse and scalar amplitudes for the $N(1520){1\over 2}^-$  
determined  from $ep \to eN\pi$ and $ep\to ep\pi^+\pi^-$ (same units as in Fig.~\ref{fig:Delta}). 
Electroproduction data are from 
CLAS~\cite{Aznauryan:2008pe,Aznauryan:2009mx,Mokeev:2012vsa,Mokeev:2015lda}.
The solid curve is the LF RQM calculations as in Fig.~\ref{fig:a12roper}.  }
\label{fig:N1520}
\end{figure}

The transition $\gamma p N(1520){3\over 2}^+$ was early on predicted in non-relativistic quark models (nr QM) 
to exhibit a radical change of its helicity structure from dominantly $A_{3/2}$ at $Q^2 = 0$ to dominant $A_{1/2}$
with increasing $Q^2$. The verification of this prediction was an early triumph of the quark model and demonstrated
that the quark model is not just applicable to baryon spectroscopy but also to the study of the internal structure of  
baryon states.  Figure~\ref{fig:N1520} shows all three electrocoupling amplitudes. $A_{3/2}$ is clearly dominant over 
$A_{1/2}$ at the photon point, while at $Q^2 > 0.5$GeV$^2$ $A_{1/2}$ is larger than $A_{3/2}$, and at 
$Q^2 > 2$GeV$^2$ $A_{1/2}$ is clearly the dominant amplitude.        
\begin{figure}[tbh]

\hspace{1.0cm}\includegraphics[height=6.0cm,width=6.0cm]{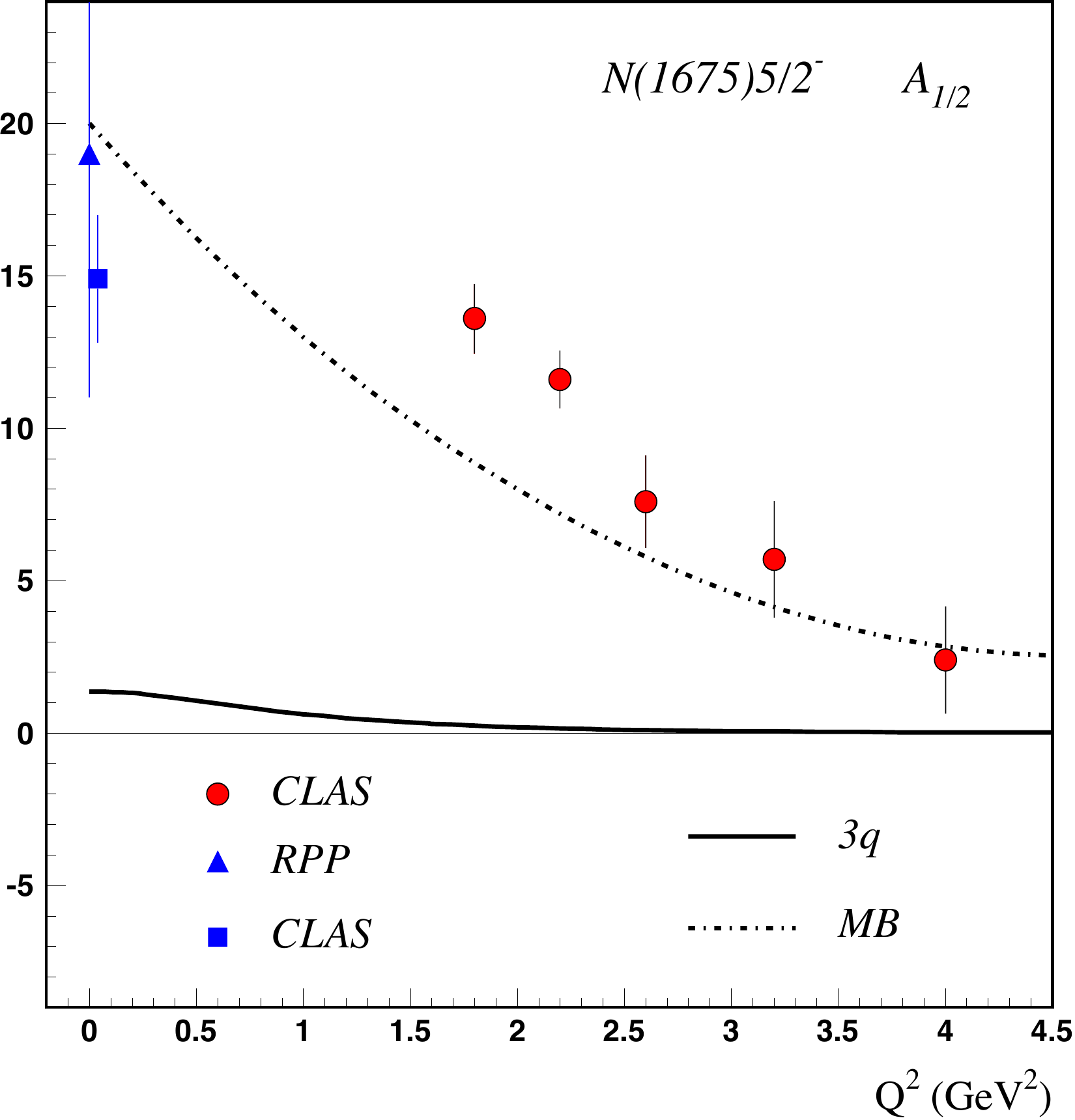}

\hspace{1.0cm}\includegraphics[height=6.0cm,width=6.0cm]{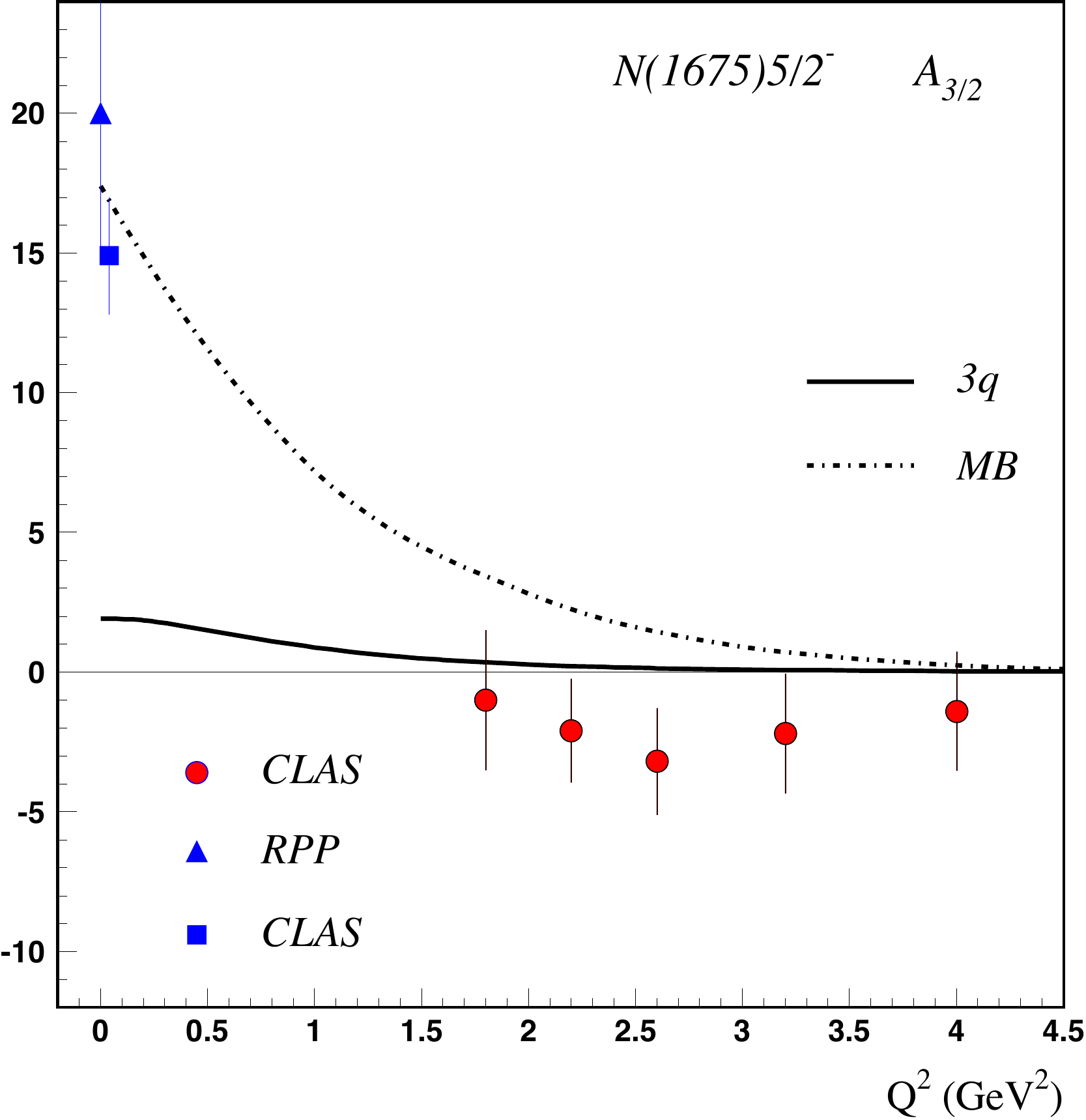}
\caption{The transverse amplitudes $A_{1/2}$ (top) and $A_{3/2}$ (bottom) of the $N(1675){5\over 2}^+$ 
have been determined in $ep \to e\pi^+n$ (same units as in Fig.~\ref{fig:Delta}). $A_{1/2}$ has a significantly non-zero magnitude dominated by
meson-baryon contributions, while within the systematic uncertainties $A_{3/2}$ is consistent with zero at 
$Q^2 > 1.7$GeV$^2$. Quark contributions are very small in both cases. Data from Ref.~\cite{Park:2014yea,Aznauryan:2014xea}.}
\label{fig:N1675}
\end{figure}
\subsection{Baryon states in the 1.7 GeV mass range} 
\label{N1675}
Differential electroproduction cross sections of $ep\to e\pi^+n$ have recently been published in the mass 
range from 1.6 to 2.0 GeV~\cite{Park:2014yea}. 
The so-called third resonance region is the domain of several nearly mass degenerate states with masses near 1.7~GeV. 
Several states, e.g. $N(1675){5\over 2}^-$ and $N(1650){1\over 2}^-$ belong to 
the $[70,1^-]_1$ supermultiplet, while the $N(1680){5\over 2}^+$ quark state is assigned to $[56, 2^+]_2$.
 Depending on the multiplet assignment, we may 
expect quite different strengths and $Q^2$ dependences of the quark components. 
For example, the quark structure of the $N(1675){5\over 2}^-$ leads to a suppressed 3-quark transition amplitude 
from the proton, i.e. $A^q_{1/2} = A^q_{3/2} = 0$. This suppression of the quark components has been verified 
for the $N(1675){5\over 2}^-$ in explicit quark model calculations. 
 We employed this suppression to directly access the non-quark components~\cite{Aznauryan:2014xea}. 
 The results are shown in  Fig.~\ref{fig:N1675}. The constituent quark model predictions are from 
Ref.~\cite{Santopinto:2012nq}. Shown predictions for the meson-baryon (MB) contributions are absolute values of 
the results from the dynamical coupled-channel model (DCCM)~\cite{JuliaDiaz:2007fa}.They are in qualitative 
agreement with the amplitudes extracted from experimental data, i.e. have considerable coupling through the $A_{1/2}$ 
amplitude and much smaller $A_{3/2}$ amplitude at $Q^2 \ge 1.8$~GeV$^2$. 

\begin{figure}[ht]
\hspace{-0.2cm}\includegraphics[height=3.5cm,width=8.5cm]{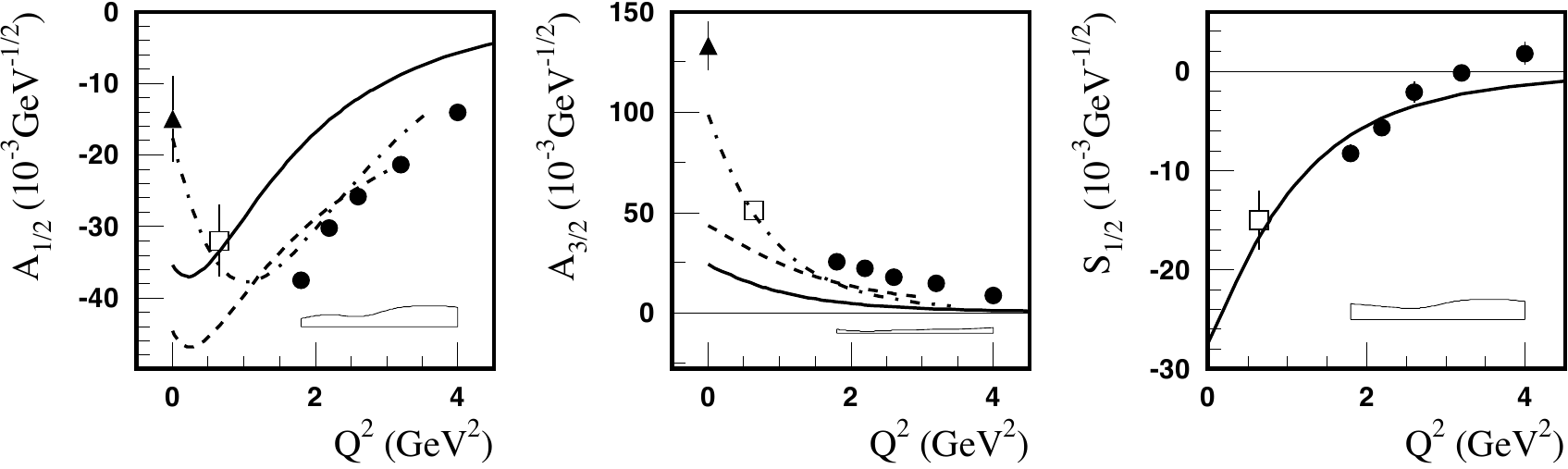}

\caption{\small The transverse amplitudes and the scalar amplitude of the $N(1680){5\over 2}^-$ have been 
determined in $ep \to e\pi^+n$ (data from Ref.~\cite{Park:2014yea}). The curves correspond to quark model calculations: dahed:~\cite{Merten:2003iy}, solid:~\cite{Santopinto:2012nq}, and dashed-dotted:~\cite{Close:1989aj}. The open boxes at the bottom indicate the model dependencies.} 
\label{fig:N1680}
\end{figure}

Figure~\ref{fig:N1680} shows the results for 
the $N(1680){5\over 2}^+$ resonant state~\cite{Park:2014yea}.  
There is  a rapid drop with $Q^2$ of the $A_{3/2}$ amplitude, which 
dominates at $Q^2=0$, while the $A_{1/2}$ amplitude, which at $Q^2=0$ makes a minor contribution,  
becomes the leading amplitude at larger $Q^2$. This change of the helicity structure is expected, 
but it is less rapid than predicted by quark models. Further studies are required to come to more definitive 
conclusions on such behavior. 

\section{Summary \& Outlook}
\label{summary}
In section~\ref{intro} we referred to the dramatic transitions that occurred in the microsecond old universe. 
The energy available at the CEBAF electron accelerator as well as at ELSA and MAMI are well matched to study 
the details of this transition in 
searching for "missing" baryon states and in probing the momentum-dependent light-quark mass in the $Q$-dependence of the
resonance transition amplitudes. Both of these provide insight into the phenomena of confinement, whose manifestation 
in strong QCD is still an open problem.   
    
The $N^*$ program pursued by the CLAS collaboration has made very significant contributions towards improving our 
understanding of strong QCD by charting the kinematic landscape of the excited light-quark baryon states, which led to
revealing of many new excited states. The search for new baryon states continues as much of the data the CLAS detector produced in the nucleon resonance region is still to be analyzed, so does the equally successful research on electromagnetic transition form factors of many excited states. 
The studies also led to advances in the theoretical exploration of strong QCD through 
the Dyson-Schwinger Equation based computations, through Light Cone Sum Rule calculations guided by LQCD, and 
through advances in Light Front RQM.   

The multi-channel analysis of the Bonn-Gatchina group using the open strangeness channels $\gamma p \to K^+\Lambda$ 
has revealed a series of new nucleon states in the mass range of 1.9 to 2.2 GeV. It is in this mass range where a novel 
kind of matter is predicted to emerge~\cite{Dudek:2012ag}, the hybrid baryons, or gluonic excitations, where the glue plays an active role in the excitation of 
baryon resonances.  This new kind of matter is not distinct from ordinary quark matter in terms of their $J^P$ quantum numbers, 
but due to the different intrinsic structure, the new states should exhibit $Q^2$ dependences of the electrocouplings that are 
different~\cite{Li:1991yba} from ordinary quark excitations or meson-baryon excitations. 
\begin{figure}[tbh]
\vspace{-0.5cm}\hspace{-1.0cm}\includegraphics[height=8.5cm,width=9.5cm]{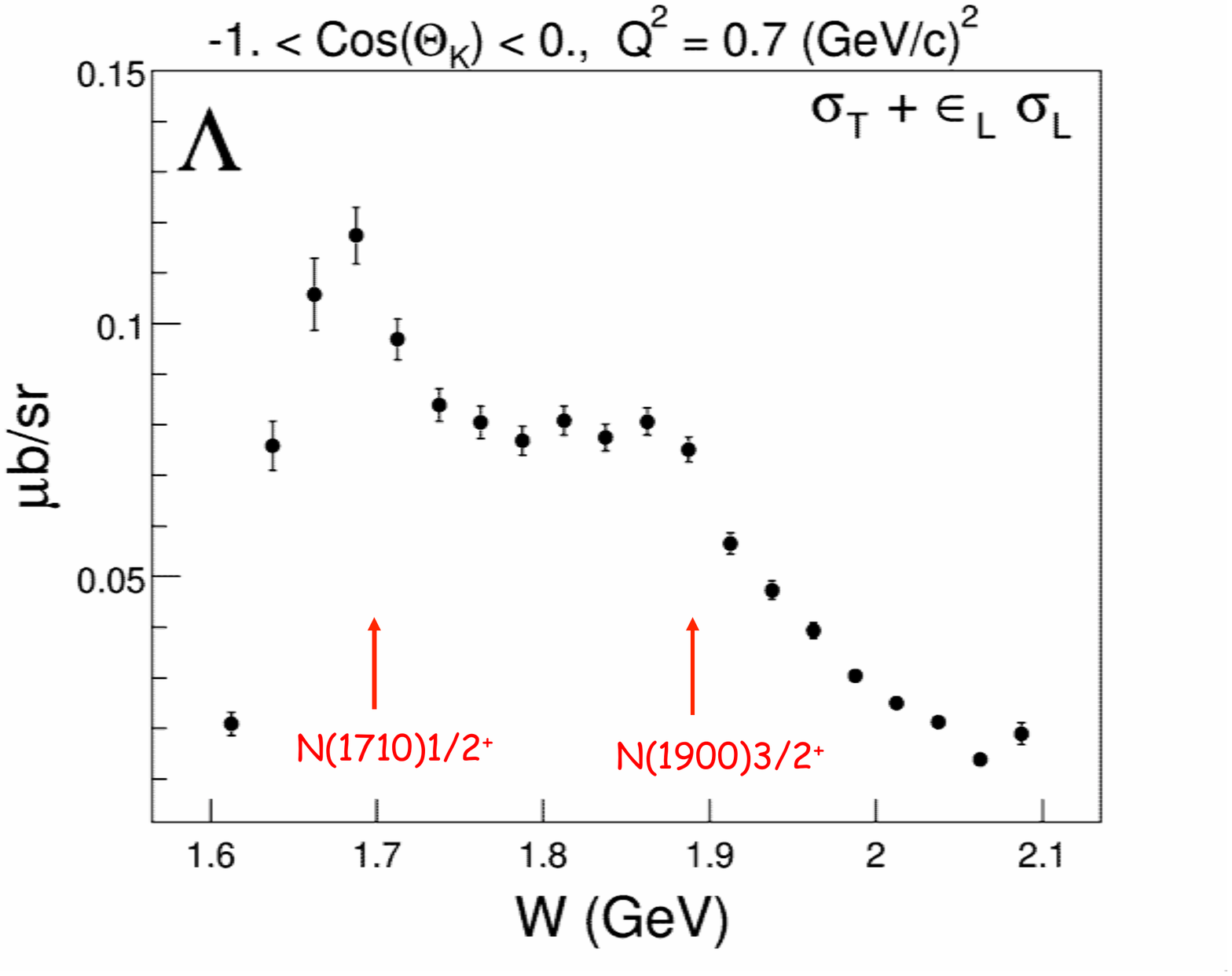}
\vspace{-1.8cm}\caption{\small Integrated cross section in the backward hemisphere of $ep \to eK^+\Lambda$. 
The shown resonances are known to couple significantly to $K\Lambda$. Other states in the same mass ranges may also contribute to the enhancements. }
\label{high_mass_states}
\end{figure} 

\begin{figure}[h]
\hspace{-0.5cm}\includegraphics[width=9.0cm,height=8.0cm,clip]{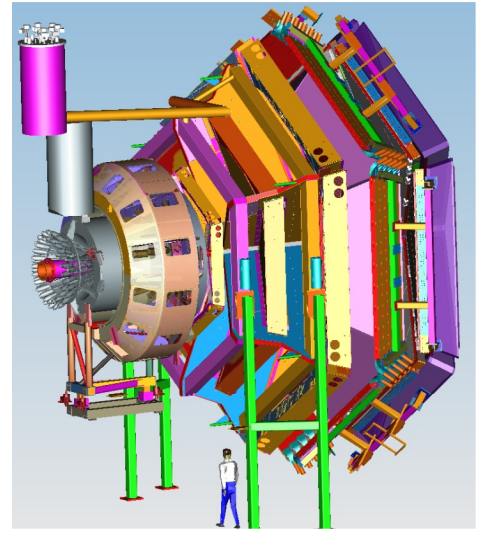}
\caption{The CLAS12 detector is projected for completion in 2017. It is based on a Torus magnet covering the forward 
angle range of $5^\circ$ to $35^\circ$, and a 5T Solenoid magnet in the central angle range of $35^\circ$ to $125^\circ$.  
New features of CLAS12 are improved particle id with FTOF for charged particle ID, improved Cherenkov counter for 
electron identification, and for $\pi$ and K separation. }
\label{clas12}      
\end{figure}

One of the goals of the new program to study the 
electrocouplings of $N^*$ states at high energies is to find this new baryon matter.  
Figure~\ref{high_mass_states} shows that in the $K^+\Lambda$ channel several of the higher mass states at 1.7~GeV and at 1.9~GeV are clearly visible in the large angle integrated cross section, and their electrocouplings may 
be studied. This requires data of much higher statistics than are currently available. Several new 
experiments have been approved to extend the $N^*$ program to higher masses and into the virtual photon dimension employing the higher energy available with the Jefferson Lab energy upgrade to 12~GeV.  
\begin{figure}[tbh]
\vspace{-0.7cm}\hspace{-0.5cm}\includegraphics[width=9.0cm,height=7.5cm,clip]{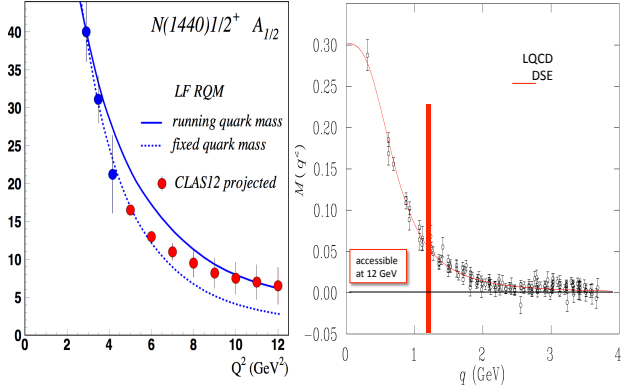}
\vspace{-2.0cm}\caption{Right panel: The quark mass versus momentum transfer. The region left to the red line is 
accessible at $Q^2 < 12$GeV$^2$, where the quark mass has dropped to about $\approx 50$~MeV. 
Left panel: The two blue lines show the estimated effect of the running quark mass 
for the Roper resonance as computed in the LF~RQM. The blue points are from measurement at beam energy of 
6~GeV~\cite{Aznauryan:2009mx}, while the red points are projections for 11~GeV.  At the highest $Q^2$ the projected value of $A_{1/2}$ for the 
running quark mass is twice as high as its value for the fixed quark mass (same units as in Fig.~\ref{fig:Delta}). }
\label{quark_mass}      
\end{figure}
The $N^*$ program at the higher energies will make use of the upgraded CLAS12 detector system that is currently 
being finalized. A cad drawing is shown in Fig.~\ref{clas12}. The higher energy allows for the exploration of a larger range in the momentum-dependent quark mass,
down to an effective dynamical mass of 50~MeV. This is shown in Fig.~\ref{quark_mass}.  For the 3-quark
system the value $q$ of the momentum transfer to a single quark is approximately given as  $q = \sqrt{Q^2}/3$. 
These studies will address the most challenging open problems in hadron physics on the nature of the dynamical 
hadron masses, the emergence of quark-gluon confinement from QCD, and its connection to dynamically generated
 symmetry breaking as the source of nearly all mass in the visible part of the universe.  

\section{Acknowledgments}
I am grateful to Inna Aznauryan, Ralf Gothe, Kijun Park and Viktor Mokeev who have contributed in many ways to the results 
discussed in this talk. I also like to thank F.-X. Girod for providing the revealing graphs of the light-cone transition charge densities, Eugene Pasyuk for a careful reading of the manuscript, and Craig Roberts for clarifying the DSE results on the resonance transition form factors.


\begin{thebibliography}{}
\bibitem{Cloet:2013jya} 
  I.~C.~Cloet and C.~D.~Roberts,
  Prog.\ Part.\ Nucl.\ Phys.\  {\bf 77}, 1 (2014)
%
\bibitem{Bazavov:2014xya} 
  A.~Bazavov {\it et al.},
  Phys.\ Rev.\ Lett.\  {\bf 113}, no. 7, 072001 (2014)

\bibitem{Bazavov:2014yba} 
  A.~Bazavov {\it et al.},
  Phys.\ Lett.\ B {\bf 737}, 210 (2014)

\bibitem{McCracken:2009ra} 
  M.~E.~McCracken {\it et al.} [CLAS Collaboration],
  Phys.\ Rev.\ C {\bf 81}, 025201 (2010)    

  \bibitem{Bradford:2006ba} 
  R.~K.~Bradford {\it et al.} [CLAS Collaboration],
  Phys.\ Rev.\ C {\bf 75}, 035205 (2007)


\bibitem{Bradford:2005pt} 
  R.~Bradford {\it et al.} [CLAS Collaboration],
  Phys.\ Rev.\ C {\bf 73}, 035202 (2006)
  
\bibitem{McNabb:2003nf} 
  J.~W.~C.~McNabb {\it et al.} [CLAS Collaboration],
  Phys.\ Rev.\ C {\bf 69}, 042201 (2004)

\bibitem{Dey:2010hh} 
  B.~Dey {\it et al.} [CLAS Collaboration],
  Phys.\ Rev.\ C {\bf 82}, 025202 (2010)

\bibitem{Anisovich:2011fc} 
  A.~Anisovich, R.~Beck, E.~Klempt, V.~Nikonov, A.~Sarantsev and U.~Thoma,
  Eur.\ Phys.\ J.\ A {\bf 48}, 15 (2012)


\bibitem{JuliaDiaz:2007fa}
B.~Julia-Diaz, T.-S.~H.~Lee, A.~Matsuyama, T.~Sato and L.~C.~Smith,
 Phys.\ Rev.\ C {\bf 77}, 045205 (2008)

\bibitem{Kamano:2013iva} 
  H.~Kamano, S.~X.~Nakamura, T.-S.~H.~Lee and T.~Sato,
  Phys.\ Rev.\ C {\bf 88}, no. 3, 035209 (2013)

\bibitem{Ronchen:2014cna} 
  D.~R\"onchen {\it et al.},
  Eur.\ Phys.\ J.\ A {\bf 50}, no. 6, 101 (2014)

\bibitem{Agashe:2014kda} 
  K.~A.~Olive {\it et al.} [Particle Data Group],
  Chin.\ Phys.\ C {\bf 38}, 090001 (2014).  

\bibitem{Paterson:2016vmc} 
  C.~A.~Paterson {\it et al.} [CLAS Collaboration],
  Phys.\ Rev.\ C {\bf 93}, no. 6, 065201 (2016)

\bibitem{Williams:2009ab} 
  M.~Williams {\it et al.} [CLAS Collaboration],
  Phys.\ Rev.\ C {\bf 80}, 065208 (2009) 

 \bibitem{Williams:2009aa} 
  M.~Williams {\it et al.} [CLAS Collaboration],
  Phys.\ Rev.\ C {\bf 80}, 065209 (2009)
 
 \bibitem{Seraydaryan:2013ija} 
  H.~Seraydaryan {\it et al.} [CLAS Collaboration],
  Phys.\ Rev.\ C {\bf 89}, no. 5, 055206 (2014)

\bibitem{Dey:2014tfa} 
  B.~Dey {\it et al.} [CLAS Collaboration],
  Phys.\ Rev.\ C {\bf 89}, no. 5, 055208 (2014)

\bibitem{Lebed:2015fpa} 
  R.~F.~Lebed,
  Phys.\ Rev.\ D {\bf 92}, no. 11, 114006 (2015)

\bibitem{Lebed:2015dca} 
  R.~F.~Lebed,
  Phys.\ Rev.\ D {\bf 92}, no. 11, 114030 (2015)

\bibitem{Aznauryan:2011qj} 
  I.~G.~Aznauryan and V.~D.~Burkert,
  Prog.\ Part.\ Nucl.\ Phys.\  {\bf 67}, 1 (2012)

\bibitem{Segovia:2014aza} 
  J.~Segovia, I.~C.~Cloet, C.~D.~Roberts and S.~M.~Schmidt,
  Few Body Syst.\  {\bf 55}, 1185 (2014)

\bibitem{Aznauryan:2015zta} 
  I.~G.~Aznauryan and V.~D.~Burkert,
  Phys.\ Rev.\ C {\bf 92}, no. 3, 035211 (2015)




\bibitem{Aznauryan:2008pe} 
  I.~G.~Aznauryan {\it et al.} [CLAS Collaboration],
  Phys.\ Rev.\ C {\bf 78}, 045209 (2008)

 \bibitem{Aznauryan:2009mx} 
  I.~G.~Aznauryan {\it et al.} [CLAS Collaboration], 
  Phys.\ Rev.\ C {\bf 80}, 055203 (2009)

\bibitem{Mokeev:2012vsa} 
  V.~I.~Mokeev {\it et al.} [CLAS Collaboration],
  Phys.\ Rev.\ C {\bf 86}, 035203 (2012)

\bibitem{Mokeev:2015lda} 
  V.~I.~Mokeev {\it et al.},
  Phys.\ Rev.\ C {\bf 93}, no. 2, 025206 (2016)

\bibitem{Segovia:2015hra} 
  J.~Segovia et al.,   
  Phys.\ Rev.\ Lett.\  {\bf 115}, no. 17, 171801 (2015)
    
\bibitem{Aznauryan:2016wwm} 
  I.~G.~Aznauryan and V.~D.~Burkert,
arXiv:1603.06692 [hep-ph].

\bibitem{Aznauryan:2012ec} 
  I.~G.~Aznauryan and V.~D.~Burkert,
  Phys.\ Rev.\ C {\bf 85}, 055202 (2012)

\bibitem{Roberts:2016dnb} 
  C.~D.~Roberts and J.~Segovia,
  arXiv:1603.02722 [nucl-th].


\bibitem{Edwards:2011jj} 
  R.~G.~Edwards, J.~J.~Dudek, D.~G.~Richards and S.~J.~Wallace,
  Phys.\ Rev.\ D {\bf 84}, 074508 (2011)
  
\bibitem{Barnes:1982fj} 
  T.~Barnes and F.~E.~Close,
  Phys.\ Lett.\ B {\bf 123}, 89 (1983).
     
\bibitem{Cano:1998wz} 
  F.~Cano and P.~Gonzalez,
  Phys.\ Lett.\ B {\bf 431}, 270 (1998)


\bibitem{Obukhovsky:2011sc} 
   I.~T.~Obukhovsky, A.~Faessler, D.~K.~Fedorov, T.~Gutsche and V.~E.~Lyubovitskij,
  Phys.\ Rev.\ D {\bf 84}, 014004 (2011)

\bibitem{Bauer:2014cqa} 
  T.~Bauer, S.~Scherer and L.~Tiator,
  Phys.\ Rev.\ C {\bf 90}, no. 1, 015201 (2014)
  
\bibitem{Suzuki:2009nj} 
  N.~Suzuki et al., 
  Phys.\ Rev.\ Lett.\  {\bf 104}, 042302 (2010)
  
\bibitem{Dudek:2012ag} 
  J.~J.~Dudek and R.~G.~Edwards,
  Phys.\ Rev.\ D {\bf 85}, 054016 (2012) 

\bibitem{Anikin:2015ita} 
  I.~V.~Anikin, V.~M.~Braun and N.~Offen,
  Phys.\ Rev.\ D {\bf 92}, no. 1, 014018 (2015)
  
\bibitem{Park:2007tn} 
  K.~Park {\it et al.} [CLAS Collaboration],
  Phys.\ Rev.\ C {\bf 77}, 015208 (2008)

\bibitem{Fedotov:2008aa} 
  G.~V.~Fedotov {\it et al.} [CLAS Collaboration],
  Phys.\ Rev.\ C {\bf 79}, 015204 (2009)
 
 \bibitem{Egiyan:2006ks} 
  H.~Egiyan {\it et al.} [CLAS Collaboration],
  Phys.\ Rev.\ C {\bf 73}, 025204 (2006) 

\bibitem{Park:2014yea} 
  K.~Park {\it et al.} [CLAS Collaboration],
  Phys.\ Rev.\ C {\bf 91}, 045203 (2015)
 
\bibitem{Li:1991yba} 
  Z.~p.~Li, V.~Burkert and Z.~j.~Li,
  Phys.\ Rev.\ D {\bf 46}, 70 (1992).


\bibitem{Tiator:2008kd} 
  L.~Tiator and M.~Vanderhaeghen,
  Phys.\ Lett.\ B {\bf 672}, 344 (2009)

\bibitem{Zou:2006tw} 
  B.~S.~Zou,
  Nucl.\ Phys.\ A {\bf 790}, 110 (2007)

\bibitem{Aznauryan:2014xea} 
  I.~G.~Aznauryan and V.~D.~Burkert,
  Phys.\ Rev.\ C {\bf 92}, no. 1, 015203 (2015)

\bibitem{Merten:2003iy} 
  D.~Merten, U.~Loring, B.~Metsch and H.~Petry,
  Eur.\ Phys.\ J.\ A {\bf 18}, 193 (2003).

\bibitem{Santopinto:2012nq} 
  E.~Santopinto and M.~M.~Giannini,
  Phys.\ Rev.\ C {\bf 86}, 065202 (2012)

\bibitem{Close:1989aj} 
  F.~E.~Close and Z.~P.~Li,
  Phys.\ Rev.\ D {\bf 42}, 2194 (1990).

\end{thebibliography}
\end{document}